\documentclass{elsart}
\journal{Nuclear Physics A}
\usepackage{amsmath,amssymb} 
\usepackage{graphicx}
\begin{document}

\begin{frontmatter}

\title{Nuclear Input for Core-collapse Models}

\author[ieec]{Gabriel Mart{\'\i}nez-Pinedo},
\author[cita]{Matthias Liebend\"orfer}, and
\author[munster]{Dieter Frekers}

\address[ieec]{ICREA and Institut d'Estudis Espacials de Catalunya,
  Universitat Aut\`onoma de Barcelona, E-08193 Bellaterra, Spain}

\address[cita]{Canadian Institute for Theoretical Astrophysics, Toronto, ON
  M5S 3H8, Canada}

\address[munster]{Institut f\"ur Kernphysik,
   Westf\"alische Wilhelms-Universit\"at M\"unster, D-48149 M\"unster,
   Germany}

\begin{abstract}
  We review the nuclear physics input necessary for the study of the
  collapse of massive stars as precursor to supernova explosions.
  Recent theoretical advances for the calculation of the relevant
  weak-interaction processes and their influence in the collapse and
  postbounce dynamics are discussed. Future improvements are expected
  to come with the advent of radioactive ion-beam facilities.
\end{abstract}

\end{frontmatter}

\section{Introduction}
\label{sec:introduction}

Stars with masses exceeding roughly 10 M$_\odot$ reach a moment in
their evolution when their iron core provides no further source of
nuclear energy generation. At this time, they collapse and, if not too
massive, bounce and explode in spectacular events known as type II or
Ib/c supernovae.  These explosions mark the formation of a neutron
star or black hole at the end of the life of the star and play a
preeminent role in the nucleosynthesis and chemical evolution of the
galaxy. The evolution in the core is determined by the competition of
gravity, that produces the collapse of the core, and the weak
interaction, that determines the rate at which electrons are captured
and the rate at which neutrinos are trapped during the collapse.

The early phases, known as presupernova evolution, follow the
late-stage stellar evolution until core densities just below
$10^{10}$~g~cm$^{-3}$ and temperatures between 5 and 10 GK are
reached. Stellar evolution until this time requires the consideration
of an extensive nuclear network, but is simplified by the fact that
neutrinos need only be treated as a sink of energy and lepton number.
This is no longer valid at later stages of the collapse: as the weak
interaction rates increase with the increasing density, the neutrino
mean free paths become shorter so that the neutrinos eventually
proceed through all phases of free streaming, diffusion, and trapping.
An adequate handling of the transitions between these transport
regimes necessitates a detailed time- and space-dependent bookkeeping
of the neutrino distributions in the core. During collapse, only the
electron neutrinos, \( \nu_e \), are important because the positron
abundance, which would lead to electron antineutrino emission by
capture on neutrons, is very low under electron-degenerate conditions.
Later in the evolution the electron degeneracy is partially lifted and
in addition to the electron flavor neutrinos also heavy neutrinos, \(
\nu_{\mu} \), \( \nu_{\tau} \), and their antiparticles, are usually
included in numerical simulations of core collapse and postbounce
evolution.

Advantageously, the temperature during the collapse and explosion are
high enough that the matter composition is given by nuclear
statistical equilibrium without the need of reaction networks for the
strong and electromagnetic interaction. The transition from a rather
complex global nuclear network, involving many neutron, proton and
$\alpha$ fusion reactions and their inverses, to a quasi-statistical
equilibrium, in which reactions within mini-cycles are fast enough to
bring constrained regions of the nuclear chart into equilibrium, to
finally global nuclear statistical equilibrium is extensively
discussed by~\cite{Woosley:1986}.

The crucial weak processes during the collapse and postbounce
evolution are~\cite{Bruenn:1985,Mezzacappa.Bruenn:1993b,%
Rampp.Janka:2002,Langanke.Martinez-Pinedo:2003,%
Burrows.Reddy.Thompson:2004} 
\begin{eqnarray}
    p + e^- & \rightleftarrows & n + \nu_e \label{eq:epnnu}\\
    n + e^+ & \rightleftarrows & p + \bar{\nu}_e \label{eq:enpnu}\\
    (A,Z) + e^- & \rightleftarrows & (A,Z-1) + \nu_e \label{eq:eAnuA}\\
    (A,Z) + e^+ & \rightleftarrows & (A,Z+1) + \bar{\nu}_e
    \label{eq:posA}\\ 
    \nu + N & \rightleftarrows & \nu + N \label{eq:nuNnuN} \\
    N + N & \rightleftarrows & N + N + \nu + \bar{\nu}
    \label{eq:NNNNnunu}\\ 
    \nu + (A,Z) & \rightleftarrows & \nu + (A,Z) \label{eq:nuAnuA}\\
    \nu + e^\pm & \rightleftarrows & \nu + e^\pm \label{eq:nuenue}\\
    \nu + (A,Z) & \rightleftarrows & \nu + (A,Z)^* \label{eq:nuAnuAin}\\
    e^+ + e^- & \rightleftarrows & \nu + \bar{\nu} \label{eq:eenunu}\\
    (A,Z)^* & \rightleftarrows & (A,Z) + \nu + \bar{\nu}
    \label{eq:AAnunu} \\ 
    \nu_e\bar{\nu}_e & \rightleftarrows & \nu_{\mu,\tau}
    \bar{\nu}_{\mu,\tau} 
  \label{eq:nunubar} 
\end{eqnarray}
Here, a nucleus is symbolized by its mass number $A$ and charge $Z$,
$N$ denotes either a neutron or a proton and $\nu$ represents any
neutrino or antineutrino. We note that, according to the generally
accepted collapse picture~\cite{Bethe:1990}, elastic scattering of
neutrinos on nuclei~\eqref{eq:nuAnuA} is mainly responsible for the
trapping, as it determines the diffusion time scale of the outwards
streaming neutrinos. Shortly after trapping, the neutrinos are
thermalized by energy downscattering, experienced mainly in inelastic
scattering off electrons~\eqref{eq:nuenue}.  The relevant cross
sections for these processes are collected in appendix C of Ref.
\cite{Bruenn:1985} (for an update see Ref.
\cite{Burrows.Reddy.Thompson:2004} in this volume).  Reactions
\eqref{eq:epnnu} and \eqref{eq:eAnuA} are equally important, as they
control the neutronization of the matter and, in a large portion, also
the star's energy losses. Due to their strong phase space sensitivity
($\sim E_e^5$), the electron capture rates increase rapidly during the
collapse as the density (the electron chemical potential scales like
$\sim \rho^{1/3}$) and the temperature increase. In the postbounce
phase, it is neutrino-nucleon scattering, Eq.  (\ref{eq:nuNnuN}), that
provides the main opacity, and lepton capture on nucelons, eqs.
(\ref{eq:epnnu}) and (\ref{eq:enpnu}), that are responsible for the
dominant creation and absorption of electron flavor neutrinos.  While
reactions (\ref{eq:posA}), (\ref{eq:nuAnuAin}) and (\ref{eq:AAnunu})
have not yet been routinely included in dynamical models,
nucleon-nucleon
bremsstrahlung~\cite{Hannestad.Raffelt:1998,Thompson.Burrows.Horvath:2000},
Eq. (\ref{eq:NNNNnunu}), and $\nu_e \bar{\nu}_e$
annihilation~\cite{Bond:1979,Buras.Janka.ea:2003,Keil.Raffelt.Janka:2003},
Eq. (\ref{eq:nunubar}), have replaced the traditional
electron-positron annihilation, Eq. (\ref{eq:eenunu}), in the role of
the dominant source for $\mu$ and $\tau$ neutrinos.

Numerical simulations of core collapse have a long history (see Ref.
\cite{Bethe:1990} and references therein).  After the observation of
the closeby supernova 1987A in the Magellanic Cloud, simulations in
spherical symmetry reached a climax around 1990. At that time it was
established that the neutrino transport scheme has to be performed in
separate energy groups to account for the strong (\( \propto E_{\nu}^2
\)) energy dependence of the neutrino cross sections and it was
realized that the thermalization of the neutrino distribution function
by neutrino-electron scattering plays an important role in the
deleptonization and core dynamics
\cite{Bruenn:1985,Myra.Bludman:1989,Bruenn:1989a,Bruenn:1989b,%
  Cooperstein:1993,Wilson.Mayle:1993}.  Most simulations relied on the
flux-limited diffusion equation for the neutrinos. The accuracy of
this approximation in the collapse phase was evaluated by comparison
to approaches where the full Boltzmann transport equation was solved
\cite{Mezzacappa.Bruenn:1993b}.  It took another decade until these
more sophisticated schemes also mastered the postbounce phase in
general relativistic space-time
\cite{Liebendoerfer.Mezzacappa.ea:2001,Rampp.Janka:2002}. The
narrowing technical uncertainties in spherically symmetric simulations
and the accessible information about the detailed neutrino
distribution functions sparked new interest to improve the nuclear and
weak interaction physics ingredients.  These efforts coincided with
the development of large-scale shell model calculations (see
Ref.~\cite{Caurier.Martinez-Pinedo.ea:2004} for a review) that allowed
for a more sophisticated and accurate treatment of electron capture
rates for the presupernova and collapse
phases~\cite{Langanke.Martinez-Pinedo:2003,Langanke.Martinez-Pinedo:2000,%
  Langanke.Martinez-Pinedo.ea:2003}. Moreover, the shell-model
calculations could be validated~\cite{Caurier.Langanke.ea:1999}
against the experimental data on GT distribution measured on
charge-exchange experiments. The advent of new radioactive ion beam
facilities and high resolution experimental techniques based on the
$(t,{}^3\mathrm{He})$~\cite{Sherrill.Akimune.ea:1999,Daito.Akimune.ea:1998}
and
$(d,{}^2\mathrm{He})$~\cite{Ohnuma.Hatanaka.ea:1993,Xu.Ajupova.ea:1995}
charge-exchange reactions will contribute to an improved treatment of
the electron capture rates used in simulations.

No spherically symmetric simulation of core collapse and postbounce
evolution with accurate neutrino transport produced an
explosion~\cite{Rampp.Janka:2000,%
  Mezzacappa.Liebendoerfer.ea:2001,Bruenn.DeNisco.Mezzacappa:2001,%
  Thompson.Burrows.Pinto:2003,Liebendoerfer.Rampp.ea:2005}, neither by
the prompt bounce-shock mechanism, nor by the delayed neutrino-heating
mechanism. Despite the mentioned progress in simulation technique and
input physics, it became very evident that the complicated dynamics of
the delayed supernova explosion can no longer be expected to be
captured in spherically symmetric simulations.  It has been shown that
the heating region is unstable to convective overturn which
significantly improves the efficiency of the neutrino heating behind
the stalled bounce-shock
\cite{Herant.Benz.ea:1994,Burrows.Hayes.Fryxell:1995,Janka.Mueller:1996}.
However, later investigations with two-dimensional hydrodynamics and
more accurate (particularly energy-dependent) neutrino transport still
failed to reproduce vigorous supernova
explosions~\cite{Mezzacappa.Calder.ea:1998,Buras.Rampp.ea:2003}.
Furthermore, suggestions that an enhanced neutrino luminosity due to
neutron finger convection in the protoneutron
star~\cite{Wilson.Mayle:1993} could increase the heating behind the
shock are also not confirmed by recent analytical
analysis~\cite{Bruenn.Raley.Mezzacappa:2004}. Hence, more ingredients
to a realistic supernova model must be missing or inadequately be
accounted for. There are however indications, that an additional twist
might suffice to at least produce an explosion, like for example a
very asymmetric deformation of the standing accretion shock
\cite{Blondin.Mezzacappa.DeMarino:2003,Janka.Buras.ea:2004} or effects
from magnetic fields~\cite{Thompson:2000,Akiyama.Wheeler.ea:2003,%
  Thompson.Quataert.Burrows:2004,Kotake.Sawai.ea:2004}.

In this review we will discuss nuclear physics ingredients to which
core collapse supernovae are sensitive.
The manuscript is organized as follows: first we discuss the role of
weak interactions in the presupernova evolution. Then we comment on
the main hydrodynamical features of core collapse based on a realistic
nuclear equation of state. We will see that the dynamics is governed
by entropy changes and deleptonization, which we then investigate
in detail with respect to traditional and new nuclear physics ingredients.
Finally, we discuss the influence of the recently improved collapse
physics on the post-bounce evolution.

\section{Presupernova evolution}
\label{sec:pres-evol}

\begin{figure}[htb]
  \centering
  \includegraphics[width=0.6\linewidth]{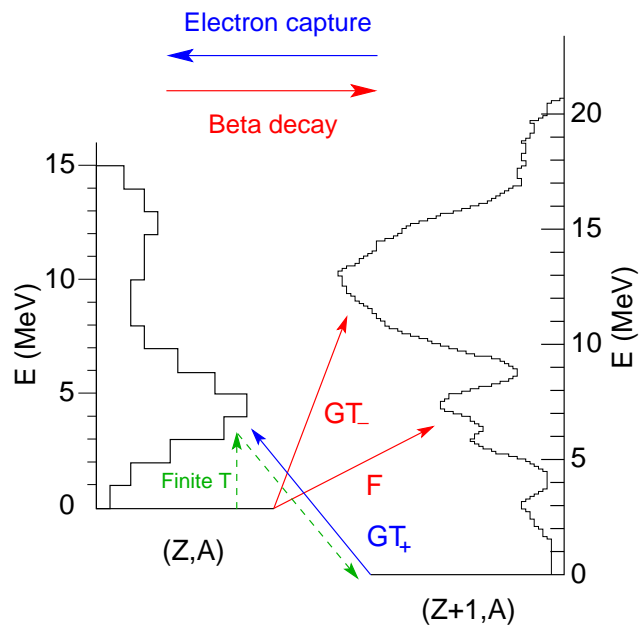}
  \caption{The figure shows schematically the electron-capture and
    beta-decay processes in the stellar environment. Electron capture
    proceeds by Gamow-Teller transitions to the $GT_+$ resonance. In
    the case of beta decay both the Fermi and Gamow-Teller resonances
    are typically outside the $Q_\beta$ window and hence not populated
    in the beta decay. Due to the finite temperature in the stellar
    environment, excited states in the decaying nucleus can be
    thermally populated. Some of these states have large GT
    transitions to low lying states in the daughter nucleus. These
    states in the decaying nucleus are called ``backresonances''
    \label{fig:scheme}}
\end{figure}

The main weak interaction processes during the final evolution of a
massive star are electron capture and beta decays. Its determination
requires the calculation of Fermi and Gamow-Teller (GT) transitions.
While the treatment of Fermi transitions (important only for beta
decays) is straightforward, a correct description of the GT
transitions is a difficult problem in nuclear structure. In the
astrophysical environment nuclei are fully ionized, so one has
continuum electron capture from the degenerate electron plasma. The
energies of the electrons are high enough to induce transitions to the
Gamow-Teller resonance. Shortly after the discovery of these
collective excitation Bethe \emph{et~al.}~\cite{Bethe.Brown.ea:1979}
recognized its importance for stellar electron capture. This process
is mainly sensitive to the location, fragmentation and total strength
of the Gamow-Teller resonance. The presence of a degenerate electron
gas blocks the phase space for the produced electron in beta decay.
Then, the decay rate of a given nuclear state is greatly reduced or
even completely blocked at high densities. However, due to the finite
temperature, excited states in the decaying nucleus can be thermally
populated. Some of these states are connected by large GT transitions
to low-lying states in the daughter nucleus, which with increased
phase space can significantly contribute to the stellar beta decay
rates. The importance of these states in the parent nucleus for the
beta decay was first recognized by Fuller, Fowler and Newman (FFN)
\cite{Fuller.Fowler.Newman:1980,%
  Fuller.Fowler.Newman:1982b,Fuller.Fowler.Newman:1982a,%
  Fuller.Fowler.Newman:1985}, who coined the term ``backresonances''
(see figure~\ref{fig:scheme}).

Over the years many calculations of weak interaction rates for
astrophysical applications have become
available~\cite{Hansen:1966,Hansen:1968,Mazurek:1973,%
  Mazurek.Truran.Cameron:1974,Takahashi.Yamada.Kondoh:1973,%
  Takahashi.ElEid.Hillebrandt:1978,Aufderheide.Fushiki.ea:1994a}. For
approximately 15 years, though, the standard in the field were the
tabulations of Fuller, Fowler and Newman
\cite{Fuller.Fowler.Newman:1980,Fuller.Fowler.Newman:1982b,%
  Fuller.Fowler.Newman:1982a,Fuller.Fowler.Newman:1985}. These authors
calculated rates for electron capture, positron capture, beta decay
and positron emission plus the associated neutrino losses for all the
astrophysical relevant nuclei ranging in mass number from 21 to 60.
Their calculations were based upon an examination of all available
experimental information in the mid 1980s for individual transitions
between ground states and low-lying excited states in the nuclei of
interest. Recognizing that this only saturated a small part of the
Gamow-Teller distribution, they added the collective strength via a
single-state representation whose position and strength was
parametrized using an independent particle model.

\begin{figure}[htb]
  \centering
  \includegraphics[width=0.6\linewidth]{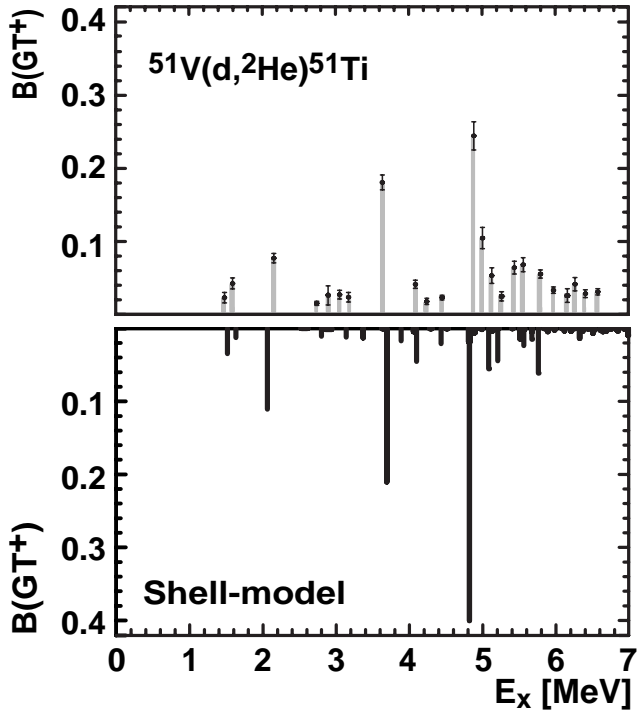}
  \caption{Comparison of the shell-model GT+ distribution (lower
    panel) for $^{51}$V with the high resolution $(d,{}^2\mathrm{He})$
    data~\cite{Baeumer.Berg.ea:2003}. The shell-model distribution
    includes a quenching factor of $(0.74)^2$.\label{fig:v51}}
\end{figure}

Recent experimental data on GT distributions in iron group
nuclei~\cite{Williams.Alford.ea:1995,El-Kateb.Jackson.ea:1994,%
  Alford.Brown.ea:1993,Alford.Helmer.ea:1990,Vetterli.Haeusser.ea:1990,%
  Rapaport.Taddeucci.ea:1983,Anderson.Lebo.ea:1990,%
  Anderson.Chittrakarn.ea:1985}, measured in charge exchange
reactions~\cite{Goodman.Goulding.ea:1980,Osterfeld:1992}, show that
the GT strength is strongly quenched (reduced), compared with the
independent-particle-model value, and fragmented over many states in
the daughter nucleus. Both effects are caused by the residual
interaction among the valence nucleons. An accurate description of
these correlations is essential for a reliable evaluation of the
stellar weak-interaction rates due to the large dependence of the
available phase-space on the electron energy, particularly for the
stellar electron-capture
rates~\cite{Fuller.Fowler.Newman:1980,Langanke.Martinez-Pinedo:2000}.
The shell-model is the only known tool to reliably describe GT
distributions in nuclei~\cite{Brown.Wildenthal:1988}. Indeed,
Ref.~\cite{Caurier.Langanke.ea:1999} demonstrated that the shell-model
reproduces very well all measured GT$_+$ distributions (in this
direction a proton is converted to a neutron, as in electron capture)
for nuclei in the iron mass range and gives a very reasonable account
of the experimentally known GT$_-$ distributions (in this direction a
neutron is converted to a proton, as in $\beta$ decay). However, the
limited experimental resolution ($\sim 1$~MeV) achieved by the
pioneering $(n,p)$-type charge-exchange experiments did not allow for
a detailed determination of the fragmentation of the GT strength in
individual states. Very recently, high-resolution GT+ distributions
measured at KVI, via the $(d,{}^2\mathrm{He})$ reaction, have become
available for two iron group nuclei,
$^{51}$V~\cite{Baeumer.Berg.ea:2003} and $^{58}$Ni
\cite{Hagemann.Berg.ea:2004}. The experimental data for $^{51}$V are
compared in figure~\ref{fig:v51} with a shell-model calculation using
the KB3G interaction~\cite{Poves.Sanchez-Solano.ea:2001}.

Several years ago, it was pointed out that the interacting shell model
is the method of choice for the calculation of stellar
weak-interaction rates~\cite{Aufderheide.Fushiki.ea:1994a,%
Aufderheide:1991,Aufderheide.Bloom.ea:1996,Aufderheide.Bloom.ea:1993a,%
Aufderheide.Bloom.ea:1993b}.  Following the work of
Ref.~\cite{Brown.Wildenthal:1988}, shell-model rates for 
all the relevant weak processes for $sd$-shell nuclei ($A=17$--39)
were calculated in Ref.~\cite{Oda.Hino.ea:1994}. This work was then
extended to heavier nuclei ($A=45$--65) based on shell-model
calculations in the complete
$pf$-shell~\cite{Langanke.Martinez-Pinedo:2000,Langanke.Martinez-Pinedo:2001}.
Following the spirit of FFN, the shell model results have been
replaced by experimental data (energy positions, transition strengths)
wherever available.

Ref.~\cite{Langanke.Martinez-Pinedo:2000} compares the shell-model
based rates with the ones computed by FFN. The shell-model rates are
nearly always smaller than the FFN ones at the relevant temperatures
and densities. The differences are caused by a reduction of the
Gamow-Teller strength (quenching) compared to the
independent-particle-model value and a systematic misplacement of the
Gamow-Teller centroid (mean energy value of the Gamow-Teller
distribution) in nuclei depending on the pairing structure. In some
cases, experimental data that were not available to Fuller, Fowler and
Newman, but could be used now, led to significant changes.

\begin{figure}[htbp]
  \includegraphics[width=0.3\linewidth]{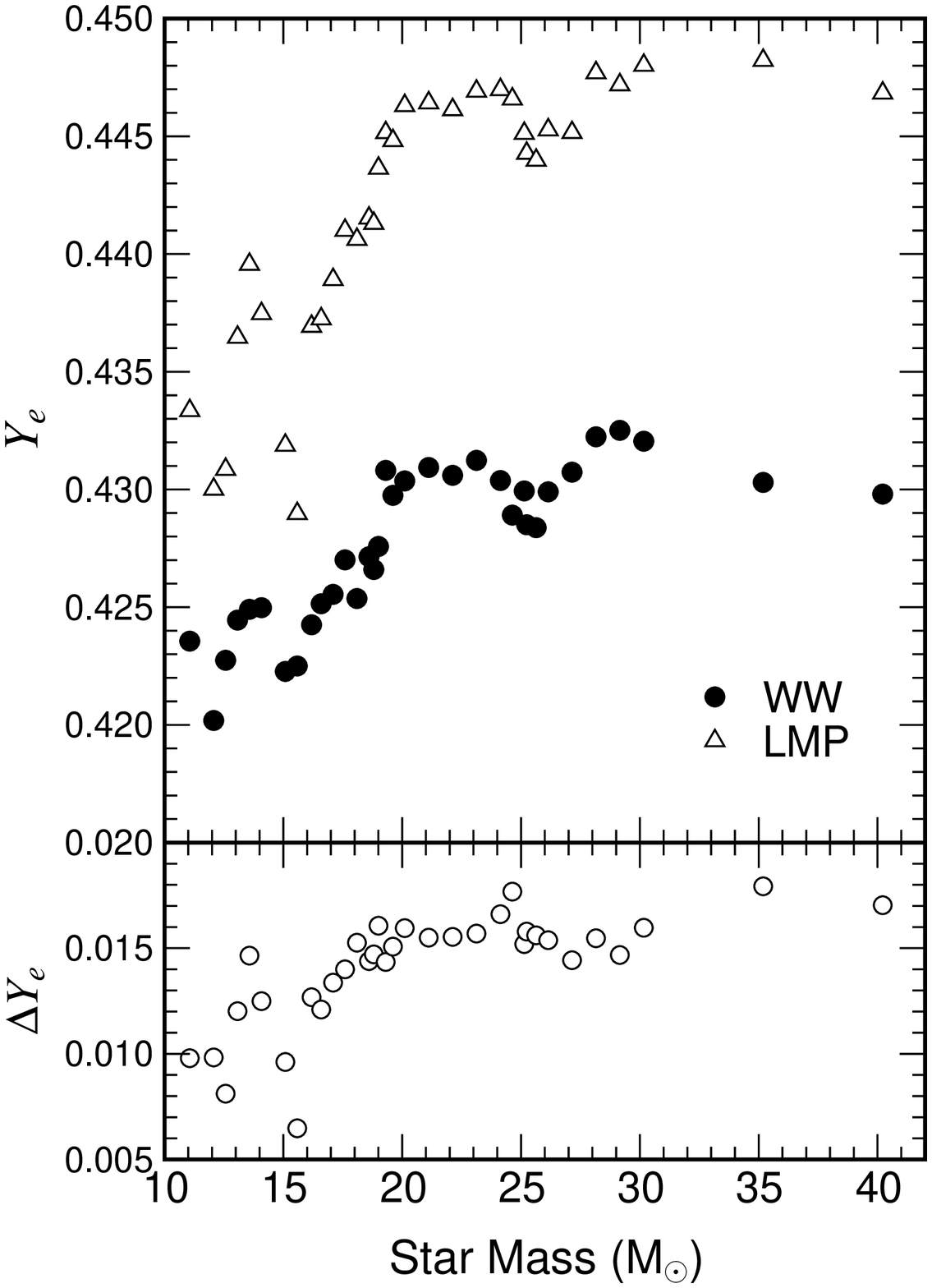}%
  \hspace{0.025\linewidth}%
  \includegraphics[width=0.3\linewidth]{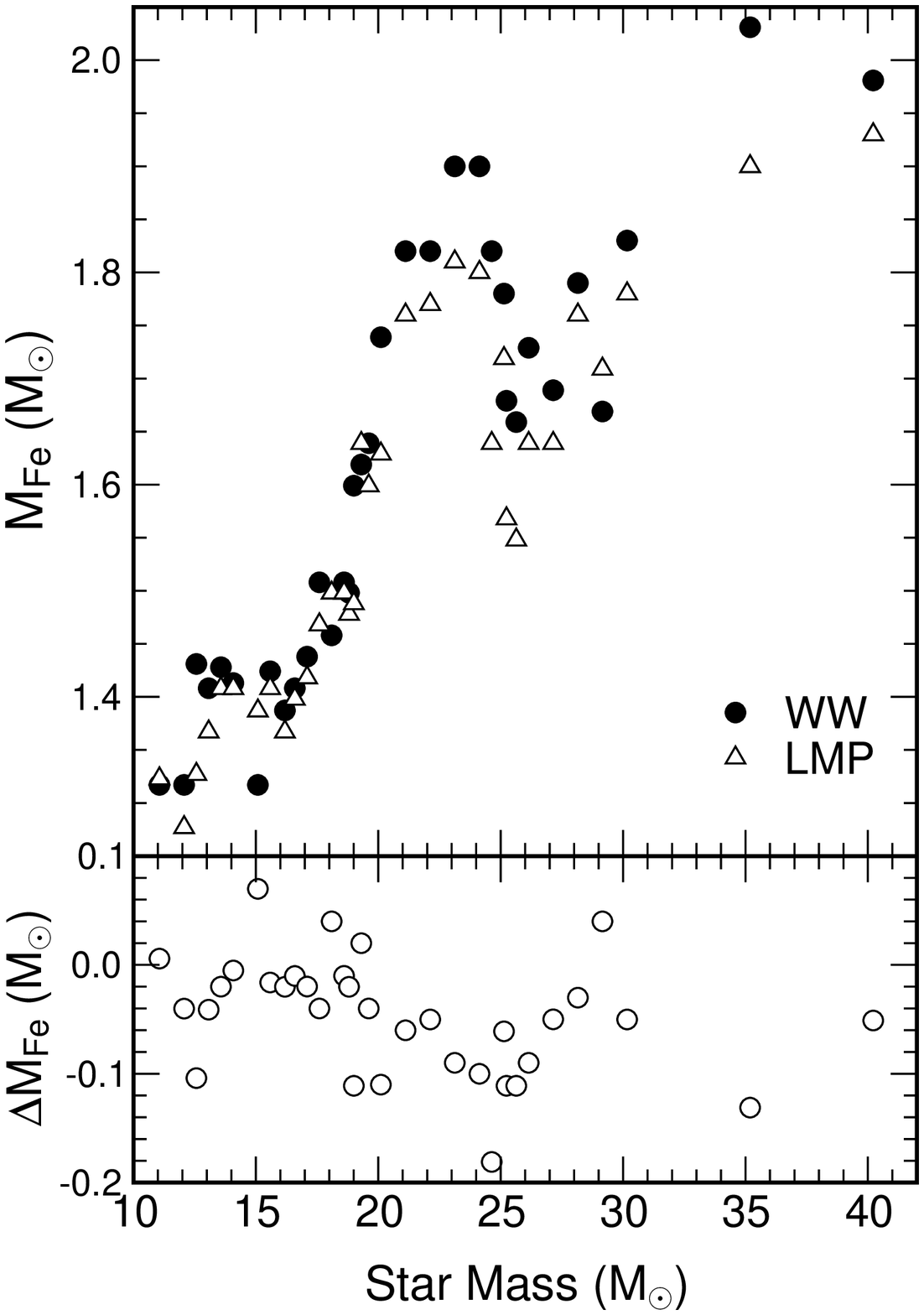}%
  \hspace{0.025\linewidth}%
  \includegraphics[width=0.3\linewidth]{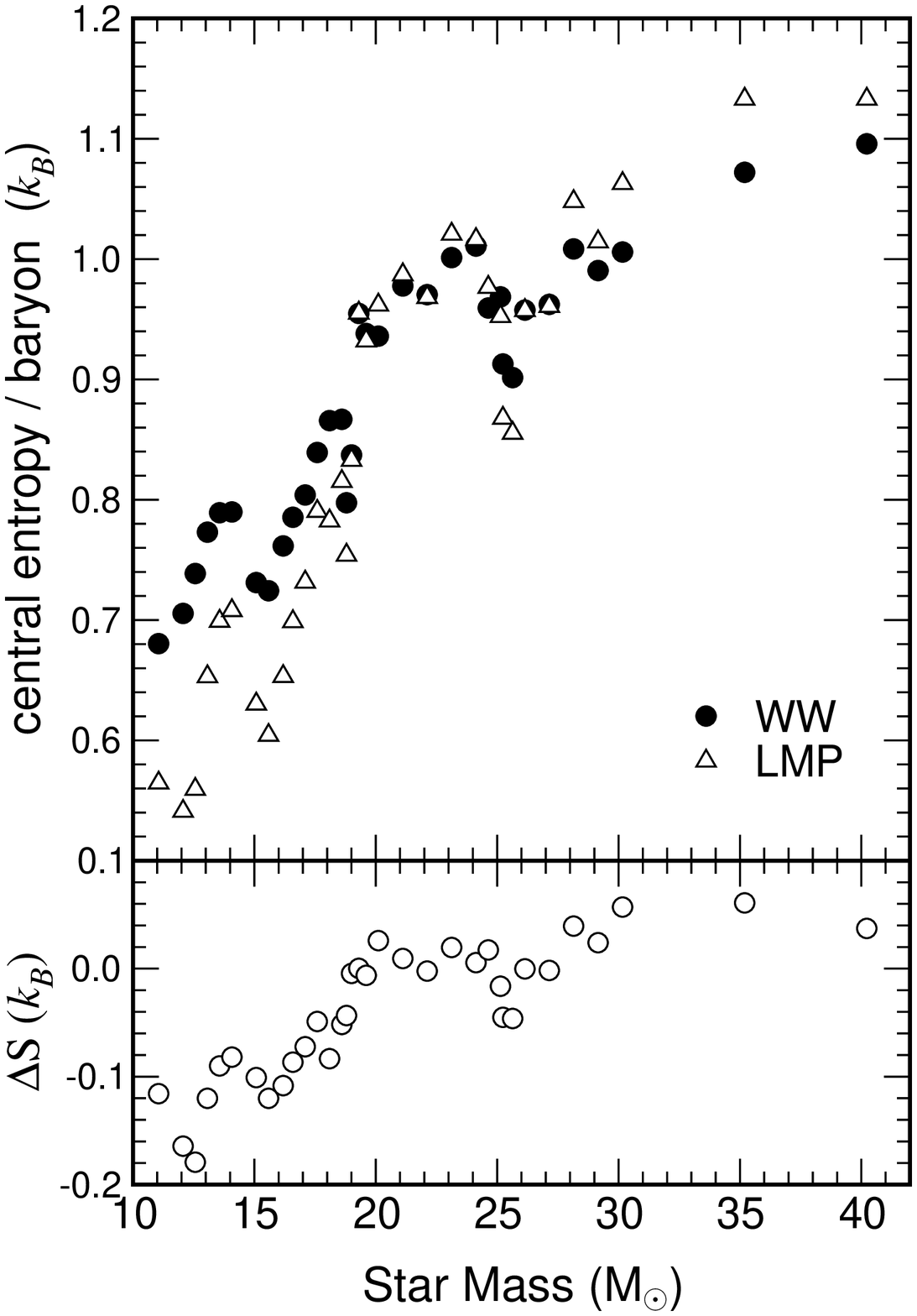}
  \caption{Comparison of the center values of $Y_e$ (left), the iron
    core sizes (middle) and the central entropy (right) for
    11--40 M$_\odot$ stars between the WW models and the ones
    using the shell model weak interaction rates (LMP)
    \cite{Heger.Langanke.ea:2001}.  The lower parts define the changes
    in the 3 quantities between the LMP and WW models.
    \label{fig:presn}}
\end{figure}

The influence of the shell-model rates in the late-stage evolution of
massive stars has been investigated in
Ref.~\cite{Heger.Langanke.ea:2001,Heger.Woosley.ea:2001} repeating the
calculations of Ref.~\cite{Woosley.Weaver:1995}, keeping the stellar
physics as close to the original studies
as possible. The new calculations incorporated the shell-model based
weak-interaction rates for nuclei with mass numbers $A=45$--65,
supplemented by the $sd$-shell nuclei rates from
Ref.~\cite{Oda.Hino.ea:1994}. The earlier calculations of Woosley and
Weaver (WW) used the FFN rates for electron capture and an older set
of beta decay rates~\cite{Mazurek:1973,Mazurek.Truran.Cameron:1974}.
Figure~\ref{fig:presn} illustrates the consequences of the shell model
weak interaction rates for presupernova models in terms of the three
decisive quantities: the central electron-to-baryon ratio $Y_e$, the
entropy, and the iron core mass.  The central values of $Y_e$ at the
onset of core collapse increased by 0.01--0.015 for the new rates.
This is a significant effect.  For example, a change from $Y_e=0.43$
in the WW model for a 20 $M_\odot$ star to $Y_e=0.445$ in the new
models increases the respective Chandrasekhar mass by about 0.075
$M_\odot$.  We note that the new models also result in lower core
entropies for stars with $M \leq 20\ M_\odot$, while for $M \geq 20\
M_\odot$, the new models actually have a slightly larger entropy.  The
iron core masses are generally smaller in the new models where the
effect is larger for more massive stars ($M \ge 20\ M_\odot$), while
for the most common supernovae ($M \le 20\ M_\odot$) the reduction is
by about 0.05~$M_\odot$.  [We define the iron core as the mass
interior to the point where the composition becomes at least $50 \%$
of iron group elements $(A \geq 48)$].  This reduction of the iron
core mass appears to be counterintuitive at first glance with respect
to the slower electron capture rates in the new models. It is,
however, related to changes in the entropy profile during silicon
shell burning which reduces the growth of the iron core just prior to
collapse \cite{Heger.Langanke.ea:2001}.

\begin{figure}[htbp]
  \includegraphics[width=0.9\linewidth]{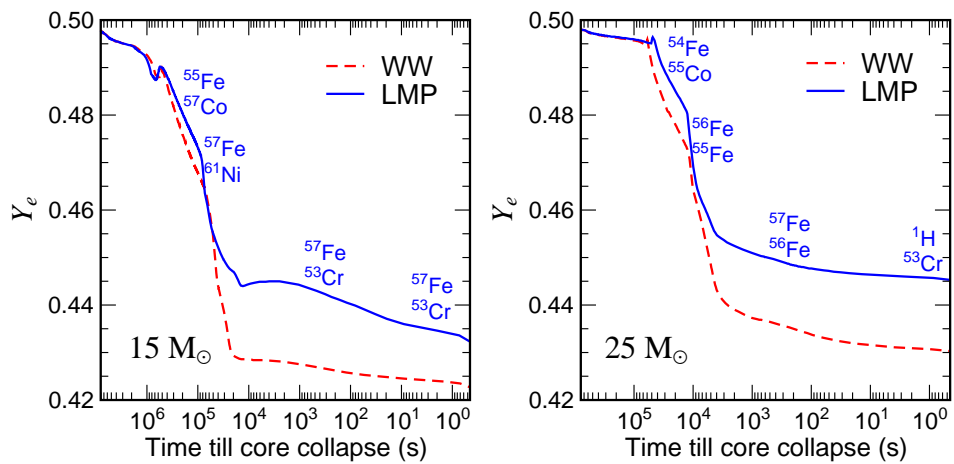}
  \caption{Evolution of the $Y_e$ value in the center of a
    15~$M_\odot$ star (left panel) and a 25~$M_\odot$ star (right
    panel) as a function of time until bounce.  The dashed line shows
    the evolution in the Woosley and Weaver models
    (WW)~\cite{Woosley.Weaver:1995}, while the solid line shows the
    results using the shell-model based weak-interaction rates of
    Langanke and Mart{\'\i}nez-Pinedo (LMP).  The most important
    nuclei in the determination of the electron-capture rate for the
    calculations adopting the shell model rates are indicated at
    different times.\label{fig:Ye}}
\end{figure}

To understand the origin of these differences it is illustrative to
investigate the role of the weak-interaction rates in greater detail.
The evolution of $Y_e$ during the presupernova phase is plotted in
figure~\ref{fig:Ye}.  Weak processes become particularly important in
reducing $Y_e$ below 0.5 after oxygen depletion ($\sim 10^7$ s and
$10^6$ s before core collapse for the 15~$M_\odot$ and 25~$M_\odot$
stars, respectively) and $Y_e$ begins a decline which becomes
precipitous during silicon burning. Initially electron capture occurs
much more rapidly than beta decay. As the shell model rates are
generally smaller than the FFN electron capture rates, the initial
reduction of $Y_e$ is smaller in the new models; the temperature in
these models is correspondingly larger as less energy is radiated away
by neutrino emission.

An important feature of the new models is shown in the left panel of
figure~\ref{fig:Ye}. For times between $10^4$ and $10^3$~s before core
collapse, $Y_e$ increases due to the fact that beta decay becomes
temporarily competitive with electron capture after silicon depletion
in the core and during silicon shell burning. This had been foreseen
in Ref.~\cite{Aufderheide.Fushiki.ea:1994b}.  The presence of an
important beta decay contribution has two effects.  Obviously it
counteracts the reduction of $Y_e$ in the core, but equally important,
beta decays are an additional neutrino source and thus they add to the
cooling of the core and a reduction in entropy.  This cooling can be
quite efficient as often the average neutrino energy in the involved
beta decays is larger than for the competing electron captures. As a
consequence the new models have significantly lower core temperatures
than the WW models after silicon burning. At later stages of the
collapse, beta decay becomes unimportant again as an increased
electron chemical potential drastically reduces the phase space.

We note that the shell model weak interaction rates predict the
presupernova evolution to proceed along a temperature-density-$Y_e$
trajectory where the weak processes are dominated by nuclei rather
close to stability. Thus it will be possible, after next generation
radioactive ion-beam facilities become operational, to further
constrain the shell model calculations by measuring relevant GT
distributions for unstable nuclei by charge-exchange reaction, where
we emphasize that the GT$_+$ distribution is also crucial for stellar
$\beta$-decays~\cite{Aufderheide.Bloom.ea:1996}.  Figure~\ref{fig:Ye}
identifies those nuclei which dominate (defined by the product of
abundance times rate) the electron capture during various stages of
the final evolution of 15~$M_\odot$ and 25~$M_\odot$ stars. An
exhaustive list of the most important nuclei for both electron capture
and beta decay during the final stages of stellar evolution for stars
of different masses is given in Ref.~\cite{Heger.Woosley.ea:2001}

In total, the weak interaction processes shift the matter composition
to smaller $Y_e$ values (see Fig.~\ref{fig:Ye}) and hence more
neutron-rich nuclei, subsequently affecting the nucleosynthesis. Its
importance for the elemental abundance distribution, however, strongly
depends on the location of the mass cut in the supernova explosion. It
is currently assumed that the remnant will have a larger baryonic mass
than the iron core, but smaller than the mass enclosed by the oxygen
shell~\cite{Woosley.Heger.Weaver:2002rmp}. As the reduction of $Y_e$
occurs mainly during silicon burning, it is essential to determine how
much of this material will be ejected.  Moreover, as show in
refs.~\cite{Froehlich.Hauser.ea:2004astro,Pruet.Woosley.ea:2004,%
Froehlich.Hauser.ea:2004}, weak interactions determine the composition
of the innermost supernova ejecta. Another important issue is the
possible long-term mixing of material during the
explosion~\cite{Kifonidis.Plewa.ea:2000}.  Changes of the elemental
abundances due to the improved weak-interaction rates are rather small
because the differences with respect to FFN occur in regions of the
star which are probably not ejected.

\section{Core-Collapse and Bounce}
\label{sec:collapse-evolution}

\subsection{Hydrodynamics}
\label{sec:hydrodynamics}

At the onset of collapse of the progenitor model, the core dynamics
becomes relevant.  The dynamics of core collapse is an interesting
example where numerical simulation stimulated the analytical
understanding. Early spherically symmetric simulations (at that time
adiabatic or based on leakage schemes)
\cite{Nadyozhin:1977,Epstein:1977,Arnett:1977,VanRiper.Arnett:1978}
suggested the separation of the collapsing material into an inner and
an outer core. The inner core fails only marginally to be pressure
supported and the sound speed stays faster than the fluid velocity at
all times. Under these conditions, the velocity develops to a linear
function of the radius. However, because the matter density and sound
speed decrease with increasing radius, there is a sonic point where
the infall velocity exceeds the sound speed so that information about
the collapsing inner core ceases to reach the outer layers.  The outer
layers therefore rather assume a free fall velocity profile in the
rarefaction of the homologously collapsing inner core.  In the
following we review different factors that determine the size of the
inner core and try to draw the connection to the initial energy
imparted to the bounce-shock.

\begin{figure}[htb]
  \begin{minipage}[t]{70mm}
    \includegraphics[width=\textwidth]{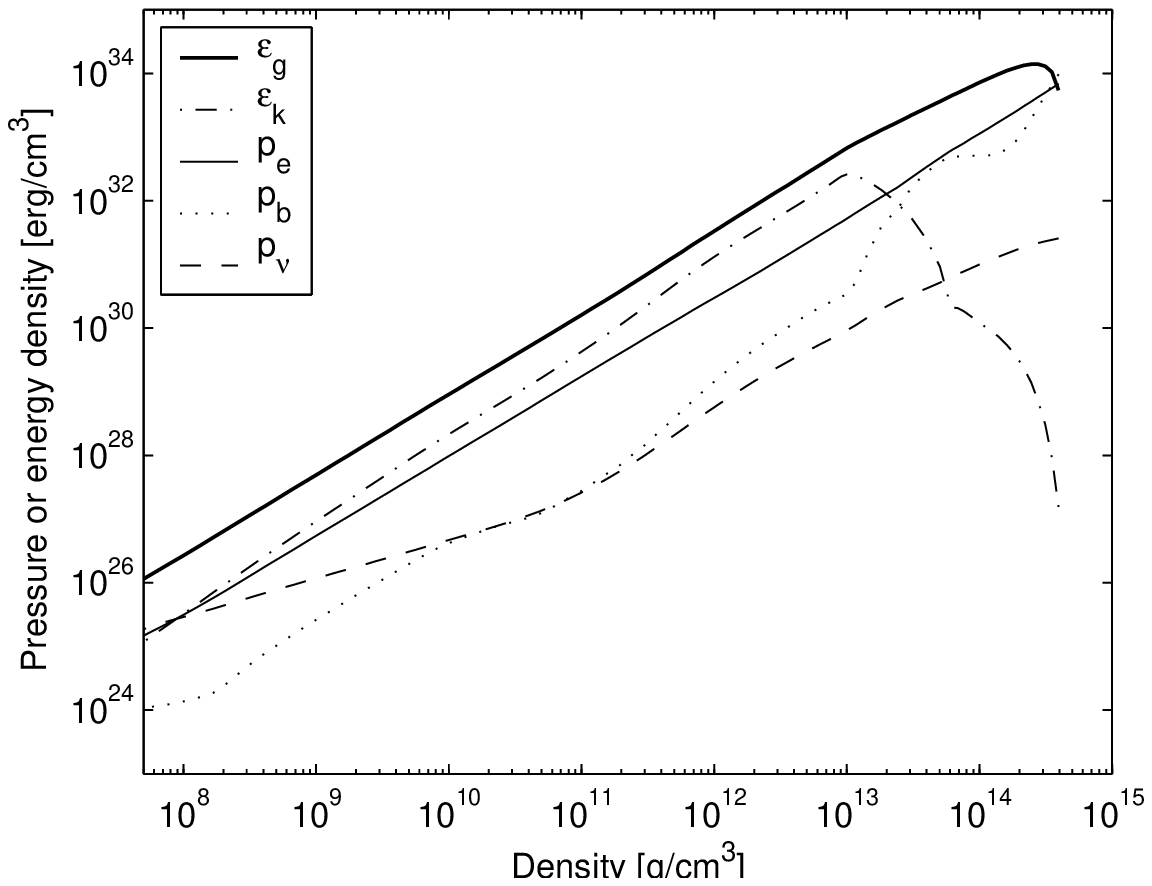}
    \caption{Energy densities as functions of rest mass density at
      bounce in model G15 \cite{Liebendoerfer.Rampp.ea:2005}. Shown
      are the gravitational energy density, \protect\( \varepsilon
      _{g}\protect \), and the kinetic energy density, \protect\(
      \varepsilon _{k}\protect \).  The total pressure splits into the
      electron pressure, \protect\( p_{e}\protect \), the baryonic
      pressure, \protect\( p_{b}\protect \), and the neutrino
      pressure, \protect\( p_{\nu }\protect \). The latter has been
      determined by taking the second angular moment of the radiation
      intensity. Note that the dynamics barely depends on the neutrino
      pressure because of the dominant electron pressure at high
      density and the negligible coupling between neutrinos and matter
      at low density \cite{Lattimer.Swesty:1991}.\label{fig:pressure.ps}}
  \end{minipage}%
  \hspace{\fill}%
  \begin{minipage}[t]{65mm}
    \includegraphics[width=\textwidth]{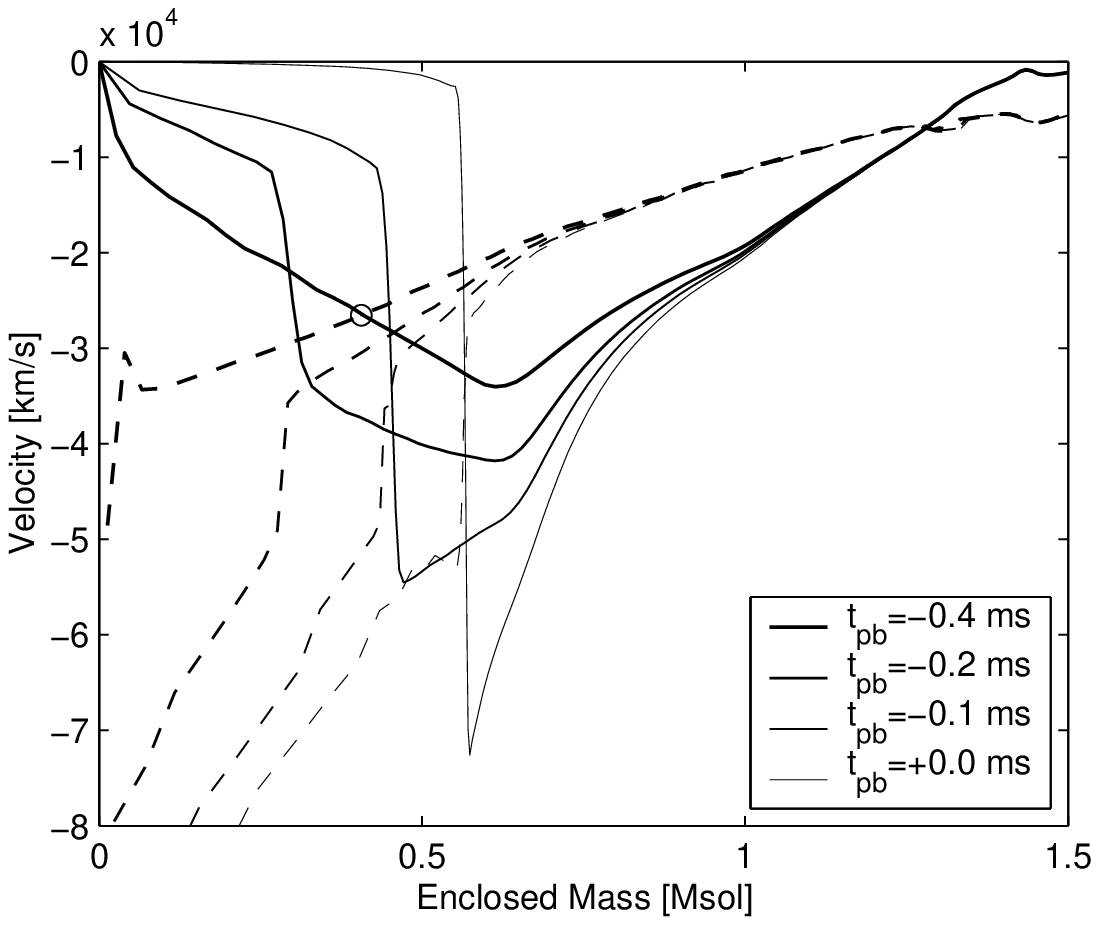}
    \caption{Velocity profiles at bounce (solid lines). The time after
      maximum central density, \protect\( t_{pb}\protect \), is
      negative before bounce. The maximum infall velocity at
      \protect\( 0.4\protect \) ms before bounce marks the edge of the
      inner core at that time. A circle is drawn at the sonic point.
      The later profiles demonstrate how a pressure wave is launched
      at the center which runs through the inner core to turn into
      a shock front close to its edge. The dashed lines represent the
      sound speed as given by the Lattimer-Swesty equation of
      state.\label{fig:velocity.ps} }
  \end{minipage}%
\end{figure}

Figure \ref{fig:pressure.ps} shows typical scales for energy
densities as a function of the rest
mass density, \( \rho  \). The data has been taken from model G15
\cite{Liebendoerfer.Rampp.ea:2005} which has been evolved with general
relativistic Boltzmann neutrino transport. A time slice
at bounce is shown. The order of magnitude of the energy density is given by
the gravitational binding energy of mass shells (thick solid line).
It is balanced by the kinetic energy (dash-dotted line) and the internal
energy. The latter determines the fluid pressure, composed from the
electron pressure (thin solid line), baryonic pressure (dotted line)
and neutrino pressure (dashed line). The dominant pressure of the
relativistic electrons scales with \( \rho ^{4/3} \) over a large
density range (in Fig. \ref{fig:pressure.ps} actually somewhat shallower
due to the radial variations in the electron fraction, \( Y_e \)).

The essentially polytropic behavior of the equation of state allows an
analytic investigation \cite{Goldreich.Weber:1980,Yahil:1983} of
collapse. It shows that the mass of the inner core, \( M_{\text{ic}} \),
is well approximated by
\begin{equation}
  \label{eq:mic}
  M_{\text{ic}}\simeq \left( \kappa /\kappa _{0}\right)^{3/2}M_{0},
\end{equation}
where \( M \) refers to the core mass and \( \kappa \)
to the coefficient in the polytropic equation of state \( p=\kappa
\rho ^{\gamma } \) with \( \gamma =4/3 \). Values with index \( 0 \)
belong to the marginally stable stage immediately before collapse.
With the coefficient for the degenerate ultra-relativistic electron
gas \cite{Shapiro.Teukolsky:1983},
\begin{equation}
  \label{eq:kappa}
  \kappa =\frac{\hbar c}{4}\left( 3\pi ^{2}\right) ^{1/3}\left(
    \frac{Y_{e}}{m_{B}}\right) ^{4/3},  
\end{equation}
one readily reproduces the conclusion that the mass of the inner core
evolves proportionally to \( Y^{2}_{e} \). The evolution
of the electron fraction is therefore the most significant ingredient
to determine the mass of the inner core at bounce. Note that the
pressure of trapped neutrinos does not reach the level of the electron
pressure because their chemical potential is lower in accordance to the
difference between the neutron and proton chemical potentials (cf.
Fig.~\ref{fig:tauegydens.ps}).

The analytical approach becomes more complex for a general adiabatic index
\( \gamma \leq 4/3 \). Nevertheless, it is possible to express the
selfsimilar solution in terms of a time-independent density profile \(
D(X) \), where \( X \) is a self-similarity variable. The inner core
mass then scales with time, \( t \), as \cite{Yahil:1983}
\begin{eqnarray}
  M_{ic} & = & \kappa ^{3/2}G^{(1-3\gamma
    )/2}(t_{\text{bounce}}-t)^{4-3\gamma }M(X)\label{eq:inner.core} \\ 
  M(X) & = & 4\pi \int _{0}^{X}dxx^{2}D(x).\nonumber
\end{eqnarray}
\( G \) is the gravitational constant, \( M(X) \) is the mass enclosed
inside the similarity variable \( X \), and \( t_{\text{bounce}} \) is
the time at bounce. The analysis demonstrates that there is a dependence
of the hydrodynamics of core collapse on the adiabatic index of the
equation of state at subnuclear densities. Due to the dominance of the
electron pressure, however, deviations from \( \gamma =4/3 \) cannot
be large.

The analytical analysis is only valid in the nonrelativistic limit.
General relativistic numerical simulations lead to a significantly
smaller mass of the inner core \cite{Takahara.Sato:1984}. Equation
(\ref{eq:inner.core}) predicts that the mass of the inner core decreases
with time if the adiabatic index falls below the critical value for
a stable core in Newtonian gravitation, i.e. \( \gamma =4/3 \). In
general relativity, the effective gravitational potential is deeper
and the critical adiabatic index for stability is larger \cite{VanRiper:1979}.
One may therefore argue that in the general relativistic
case even the \( \gamma =4/3 \) from the ultra-relativistic electron
gas falls below the critical adiabatic index. Hence, the inner core
shrinks during collapse as suggested by Eq. (\ref{eq:inner.core}).
A rigorous argument, however, would have to rely on an analytical
analysis of core collapse based on a general relativistic similarity
variable \cite{Gambino:2004}. Recent numerical simulations of core
collapse specify a reduction of about \( 20\% \) in the inner core
mass due to general relativistic
effects~\cite{Bruenn.DeNisco.Mezzacappa:2001,%
Liebendoerfer.Mezzacappa.ea:2001,Hix.Messer.ea:2003}.  
The size of the inner core is most important for the energetics of
the core-bounce:

The baryonic pressure displayed in Fig. \ref{fig:pressure.ps}
(dotted line) increases steeply at high density. Shortly above \( 10^{13} \)
g/cm\( ^{3} \) the thermal pressure component is exceeded by the
neutron degeneracy pressure. The plateau accross the phase transition
from isolated nuclei to bulk nuclear matter at \( 10^{14} \) g/cm\( ^{3} \)
is caused by the attractive nuclear forces. The phase transition with
its many possible geometrical structures and statistical fragmentation
is very interesting from a nuclear physics point of view 
\cite{Ishizuka.Ohnishi.Sumiyoshi:2003,Botvina.Mishustin:2004,%
Horowitz.Perez-Garcia.Piekarewicz:2004,Watanabe.Sato.ea:2004}.
However, a significant
impact of the microscopic phenomena in this regime on the collapse
and/or supernova dynamics remains to be demonstrated. At and above
saturation density, repulsive nuclear forces dominate the stiffness
of the equation of state. When the center of the collapsing core reaches
these densities, collapse is halted by an outgoing pressure wave.
This pressure wave travels through the inner core, where infall velocities
are subsonic, and turns into a shock wave close to its edge. The matter
in the inner core experiences a rather adiabatic pressure wave and
remains at a low entropy \( \sim 1.4 \) kB per baryon. The shock
wave in the outer core, however, heats matter to entropies larger
than \( \sim 6 \) kB per baryon so that heavy nuclei are dissociated.
If the bounce-shock were to dynamically propagate through the core
to expel outer layers in a prompt explosion, it would have to provide
the energy to dissociate the material between the edges of the inner
core and the iron core at a rate of \( 1.5\times 10^{51} \) erg per
\( 0.1 \) M\( _{\odot } \) of dissociated material (the shock additionally
suffers from unavoidable neutrino losses).

The size of the inner core does not only set the mass that will be
dissociated in the postbounce evolution, additionally it determines
the initial energy of the bounce-shock. The rebound of a larger mass
in the gravitational potential implies a larger initial energy imparted
to the outgoing shock wave. This initial shock energy has been found
to depend sensitively on the stiffness of the equation of state around
nuclear densities~\cite{VanRiper:1979,Baron.Cooperstein.Kahana:1985b,%
Takahara.Sato:1988,Bruenn:1989b,Swesty.Lattimer.Myra:1994,%
Sumiyoshi.Suzuki.ea:2004,Janka.Buras.ea:2004}.
Unfortunately, most studies used the explosion
energy of prompt hydrodynamic explosions to measure this effect. Their
results are difficult to compare because the final explosion energy
depends on the physics applied to propagate the initial shock through
the iron core.
Some apparent contradictions in the conclusions of
above-cited references suggest a few retrospective comments:

The homologous core stays very large in adiabatic calculations when
electron captures are neglected. Hence, energetic prompt explosions
are obtained. The initial shock energy increases with a decrease of
the adiabatic index in a polytropic equation of state
\cite{VanRiper:1979}.  A similar effect was found in a recent
adiabatic comparison between two realistic equations of state with
different incompressibilities at saturation density
\cite{Sumiyoshi.Suzuki.ea:2004}. The period of the lowest oscillation
mode of the inner core in the regime of the stiffened equation of
state depends strongly on the difference of its mass from the maximum
stable mass. If the mass of the inner core is much smaller than the
maximum stable mass, the oscillation period is very small with respect
to the sound crossing time in the outer part of the inner core.
Infalling material piles up around a {}``hard'' center until the
outgoing compression wave reaches the edge of the inner core, where it
turns into a relatively weak shock (see Fig. \ref{fig:velocity.ps}).
If the maximum stable mass is close to the inner core mass, the
oscillation period is larger and the oscillation is more efficiently
excited. A larger mass bounces on a {}``soft'' core and a more
energetic shock is launched at its edge. As soon as electron captures
during collapse are included, e.g. by a simple leakage scheme, the
mass of the inner core shrinks and prompt explosions are more
difficult to obtain. Only extremely soft equations of state lead to
explosions in that case \cite{Baron.Cooperstein.Kahana:1985b}. But
sufficiently soft equations of state, producing a maximum stable mass
close to the mass of the inner core, are not compatible with the
significantly larger observed neutron star masses, unless a very
extended phase transition provides softness at nuclear densities in
combination with a stiff adiabatic index in the far supranuclear
regime \cite{Takahara.Sato:1988}. Later investigations of the effect
of the equation of state at bounce were consistent with the described
trend, but ceased to produce energetic prompt explosions due to
improvements in the neutrino transport, especially after the inclusion
of neutrino electron scattering
\cite{Myra.Bludman:1989,Bruenn:1989a,Bruenn:1989b}.  Few years later,
the BCK equation of state \cite{Baron.Cooperstein.Kahana:1985a} was
replaced by the Lattimer-Swesty equation of state
\cite{Lattimer.Swesty:1991} in {}``standard'' simulations of core
collapse. Its maximum mass for the hot protoneutron star is around \(
2.4 \) M\( _{\odot } \), much larger than the mass of the inner core
\( \sim 0.5 \) M\( _{\odot } \). The consequence is a very low initial
shock energy of \( 1.9\times 10^{51} \) erg (measured in simulation
G15 \cite{Liebendoerfer.Rampp.ea:2005} according to the definition
used in Ref. \cite{Bruenn:1989b}) instead of the 5--$10\times
10^{51}$ erg obtained in previous studies with the BCK equation of
state \cite{Bruenn:1989b}. The dynamical shock stalls already at \( 3
\) ms after bounce.
The ensuing expansion of the accretion front does
not depend on the exhausted shock energy anymore. As we will
briefly discuss in section \ref{sec:postbounce}, the expansion
is determined by the accretion rate in the gravitational
potential and the deleptonization rate of the accumulated matter.
This explains why a comparison of
different choices of incompressibilities at saturation density in Ref.
\cite{Swesty.Lattimer.Myra:1994} showed no discernible consequences
for the shock propagation on a time scale of \( 50 \) ms. The study
did not resolve the few milliseconds after bounce where transient
effects of the compressibility variations may manifest in the energy
of the short-lived bounce-shock. An update on the comparison of three
different equations of state on the longer time scale of postbounce
simulations with accurate neutrino transport is given in Ref.
\cite{Janka.Buras.ea:2004}.

In summary, the crucial element in core collapse calculations is the
determination of the mass of the inner core because it determines the
point of shock formation at bounce and the initial energy that is
imparted to the shock. The following
paragraphs focus on the weak interactions which directly or indirectly
determine the core mass through the evolution of electron fraction and
entropy.

\subsection{Deleptonization}
\label{sec:deleptonization}

The deleptonization and entropy changes during collapse are determined
by the interplay between three different physical processes: (i) the
transition of protons to neutrons by electron capture and neutrino
emission (for example by reactions (\ref{eq:epnnu}) or (\ref{eq:eAnuA})),
(ii) isoenergetic neutrino scattering (e.g. by reactions
(\ref{eq:nuNnuN}) or (\ref{eq:nuAnuA})),
and (iii) neutrino thermalization (e.g. by reactions
(\ref{eq:nuenue}) or (\ref{eq:nuAnuAin})).
Each of these processes is rather
straightforward to describe in a gas of free nucleons.  In reality,
however, the by far most abundant nuclear species in the cold
collapsing matter are neutron-rich heavy nuclei (see
Fig.~\ref{fig:abundances}). While the described weak interactions
basically stay the same, they rather involve nucleons bound in nuclei
than free ones. This leads to the interesting expectation that the
nuclear structure is probed in many nuclei that are difficult or
impossible to explore under terrestrial conditions. In the following
we first describe the principle importance of each of above processes for the
deleptonization of the core before we get to their detailed
quantitative description and dependence on nuclear structure in section
\ref{sec:electr-capt-during}.

\begin{figure}
  \includegraphics[width=0.5\textwidth]{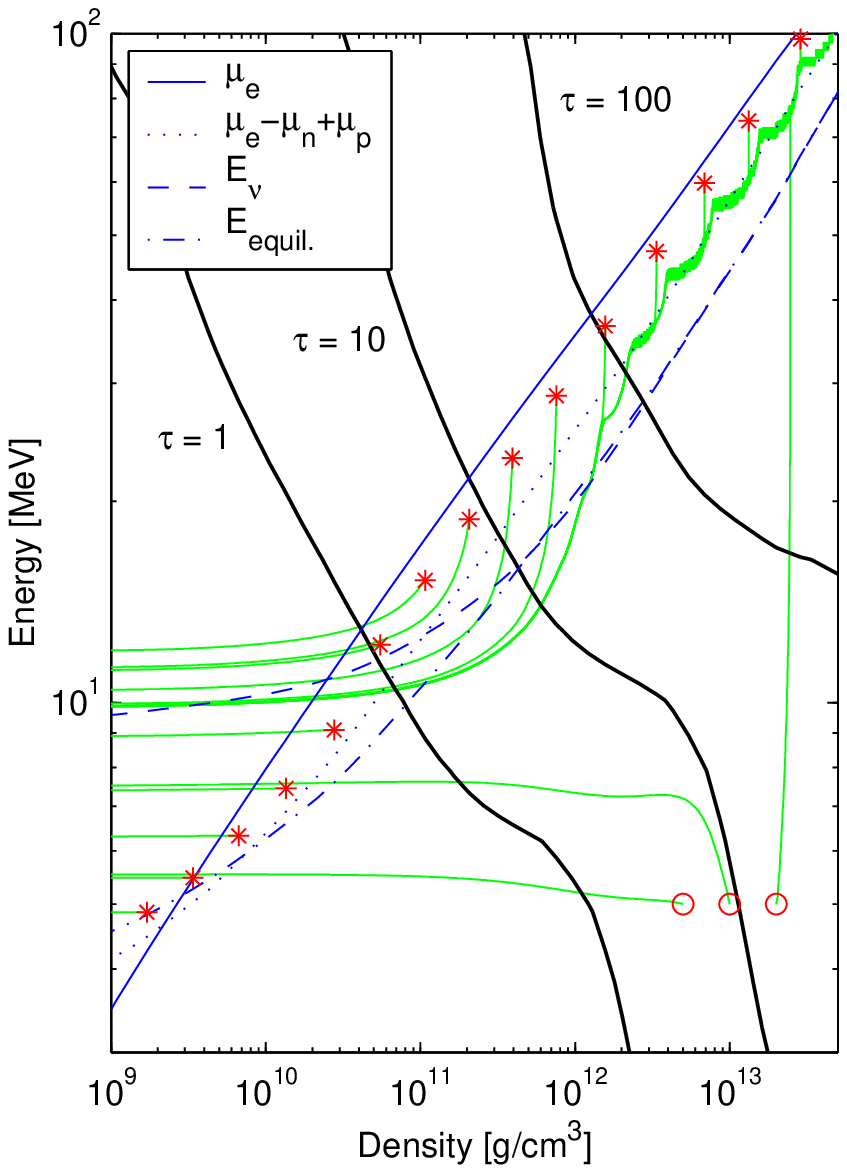}%
  \hspace{\fill}
  \includegraphics[width=0.5\textwidth]{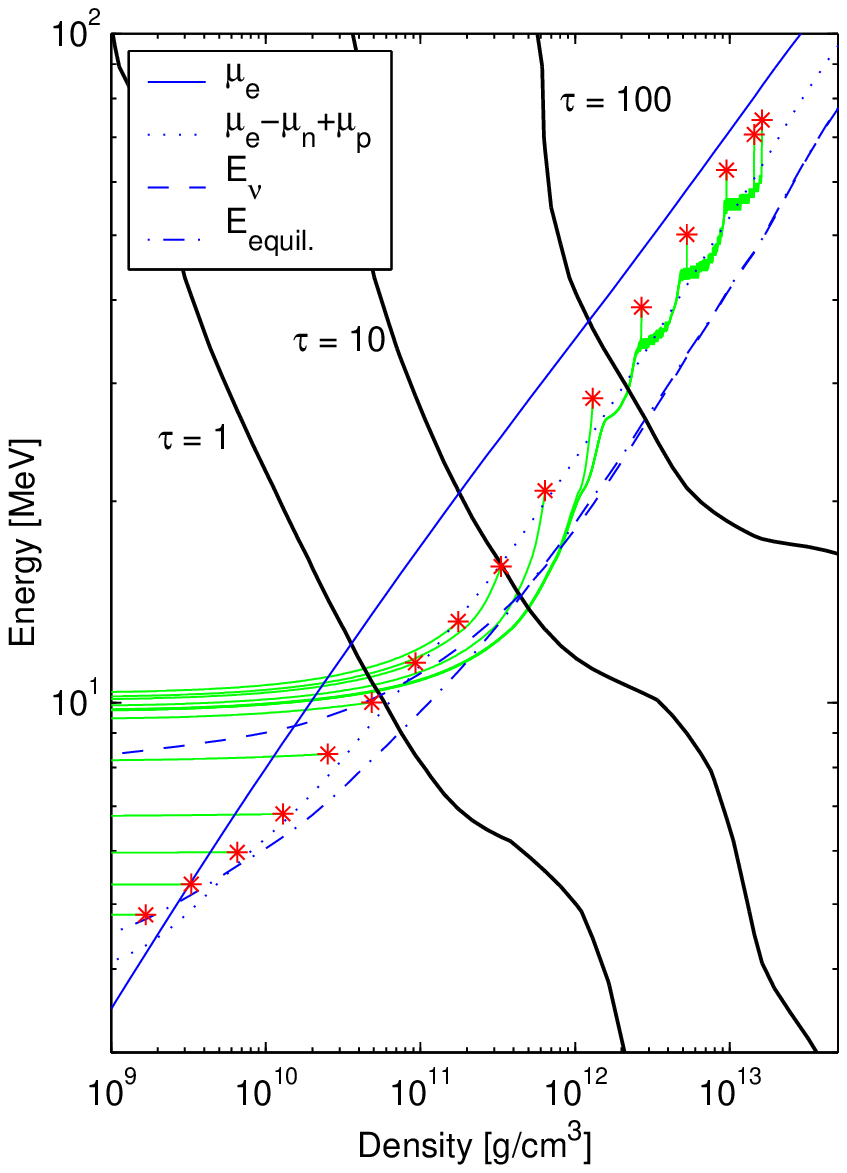}
  \caption{Shown are the electron chemical potential, \protect\( \mu
  _{e}\protect \) (straight solid line), and its distance to the
  difference between the nucleon chemical potentials, \protect\( \mu
  _{e}-\mu _{n}+\mu _{p}\protect \) (dotted line). This quantity corresponds
  to the electron neutrino chemical potential where the neutrinos
  are in equilibrium with matter (recognizable in the figure as the
  domain where the mean neutrino energy, \( E_{\nu} \), is equal to
  the equilibrium energy, \( E_{\text{equil.}} \)). The asterisks represent the
  average energy of neutrinos produced by electron capture. The
  neutrinos escape in a two-dimensional diffusion process in energy
  and space. Thick solid lines connect locations of optical depths
  \protect\( \tau =1\protect \), \protect\( \tau =10\protect \), and
  \protect\( \tau =100\protect \).
  For each emitted average energy, a thin solid line indicates the
  trajectory of fastest escape as a function of neutrino energy
  and density.
  At subnuclear densities (e.g. circle at \protect\( \rho =5\times
  10^{12}\protect \) g/cm\protect\( ^{3}\protect \)), low energy
  neutrinos start to escape and to free phase space for the
  downscattering of higher energy neutrinos around \protect\( \rho
  =10^{11}-10^{12}\protect \) g/cm\protect\( ^{3}\protect \).  In this
  region, the figure illustrates the nontrivial interplay between
  neutrino emission, thermalization, and diffusion. The left panel 
  corresponds to a time slice at bounce in model G15
  \cite{Liebendoerfer.Rampp.ea:2005} 
  with conventional input physics. The right panel corresponds to
  a simple test calculation where, in the spirit of
  Ref. \cite{Messer.Liebendoerfer.ea:2003}, the $f_{5/2}$ orbit in the
  Bruenn scheme was unblocked to illustrate the discussion of the
  improved electron capture rates in section \ref{sec:electr-capt-during}.
  \label{fig:tauegydens.ps}}
\end{figure}

As the density, \( \rho \), increases, the electron chemical potential
increases with \( \mu _{e}\propto \rho ^{1/3} \) \cite{Bethe:1990},
see also Fig. \ref{fig:tauegydens.ps} (thin solid line). The electrons
fill higher energy levels and electron captures on free or bound
protons become more and more likely. As long as the density is lower
than \( 5\times 10^{10} \) g/cm\( ^{3} \), the neutrinos escape freely
and the deleptonization rate is determined by the electron capture
rate.  Figure \ref{fig:ecrates} compares the electron capture reaction
rates on free protons with the electron capture reaction rates on
nuclei. The former is by an order of magnitude larger than the latter.
But the nuclei are by orders of magnitude more abundant targets.  It
is difficult to decide from first principles whether electron captures
on free protons or nuclei would dominate in core collapse (see for
example
\cite{Bethe.Brown.ea:1979,Fuller:1982,Cooperstein.Wambach:1984}):
Because the free proton fraction is very sensitive to small variations
in the \( Y_e \) value, it can be huge if the \( Y_e \) value is
somewhat larger than a reference value or it can be negligible if the
\( Y_e \) value is somewhat smaller than a reference value.  This had
interesting consequences in core collapse simulations. As described
above, progenitor models for different main sequence masses or with
different sophistication of the included weak interaction physics may
run into collapse with quite different electron fraction profiles.
Collapse simulations with Boltzmann neutrino transport, however,
showed that most of the initial variations in the inner core structure
converged to the same trajectories during core collapse, leading to
very similar positions of shock formation and nearly identical
neutrino bursts \cite{Messer:2000,Liebendoerfer.Messer.ea:2002}.
These simulations used the standard treatment of electron
capures~\cite{Bruenn:1985} that inhibited the dominant electron
capture on nuclei as soon as nuclei exceeded neutron number $N=40$.
This leads to a more shallow deleptonization slope at a density \(
\sim 2\times 10^{10} \) g/cm\(^3\) as illustrated in the left panel of
Fig. \ref{fig:evol}.  But the deleptonization is not slowed down for
long because a high electron fraction at increasing density (where
nuclei become more neutron rich) implies a steeply rising free proton
fraction until the deleptonization resumes on a similarly effective
time scale by captures on free protons. The \( Y_e \) value at which
this happens is determined by the intrinsic properties of the nuclei
at a given density and temperature and therefore independent from the
initial electron fraction or proton abundance. Hence, the
deleptonization of different models converges toward a ``norm''
electron fraction trajectory. A convergence of this kind will always
take place when the product of electron capture reaction rate and
target abundance dominates for a nucleus whose abundance strongly
depends on the electron fraction.

This appears somewhat contradictory to results of earlier extensive
studies of the effect of the free proton fraction on core collapse.
Significant differences in the deleptonization and inner core size
at bounce have been reported \cite{Bruenn:1989a}. This result becomes
understandable if one considers that the investigation has not been
performed with different \emph{initial} free proton mass fractions.
Instead, different free proton mass fractions have been set by variations
of the asymmetry energy in the equation of state. A smaller asymmetry
energy lowers the electron fraction at which the free proton fraction
becomes significant enough to perform electron captures effectively
on the infall time scale. It therefore leads to a lower ``norm'' trajectory for
the electron fraction and to a smaller inner core at bounce. However,
if the asymmetry energy is varied together with the surface energy
to reproduce experimental constraints, the configuration of the nuclei
is less significantly altered and similar ``norm'' trajectories
are produced with negligible consequences for core bounce
\cite{Swesty.Lattimer.Myra:1994}.

We therefore conclude that the tight feedback between electron fractions
and free proton abundances defines a limiting {}``norm'' trajectory
for the deleptonization during infall. It is determined by the mass
formula of nuclei according to its implementation in the equation of
state. The deleptonization up to neutrino trapping can occur
faster than {}``norm'' by the availability of additional channels
for electron capture, but hardly slower. The free proton mass fraction
is not a relevant independent input quantity of the progenitor structure,
it evolves rather symptomatically: in models that proceed along the
{}``norm'' trajectory, it is higher; in models that deleptonize faster,
it is smaller. For example the most recent simulations with
sophisticated nuclear input physics predict a deleptonization that is
faster than the {}``norm'' because captures on nuclei provide the most
relevant electron capture channel throughout core
collapse~\cite{Langanke.Martinez-Pinedo.ea:2003,Hix.Messer.ea:2003}.
As we will discuss in more detail in section
\ref{sec:electr-capt-during}, the entropy also becomes slightly
smaller. Both effects contribute to reduce the proton abundances when
treated consistently with the hydrodynamical evolution so that
electron capture on free protons becomes an almost inactive channel.

The large electron capture rates at densities higher than \( 5\times
10^{12} \) g/cm\( ^{3} \) are not relevant anymore because the
neutrino scattering opacities get so large that the time scale for
neutrinos to leave the star becomes larger than the production time
scale. The neutrino phase space fills up and further electron captures
are blocked for neutrino emission below the neutrino Fermi energy. In
this regime the state of matter is best described by the density,
entropy, and lepton fraction, \( Y_{\ell } \). The lepton fraction
becomes the independent variable that slowly changes on a diffusion
time scale while the electron fraction, \( Y_{e} \), almost
immediately represents the local equilibrium state between the fluid
and the trapped neutrino radiation field. The equilibrium is
established by neutrino-electron scattering. Due to the small electron
mass, the energy transfer between neutrino and the relativistic
electrons is always inelastic and efficient while the much more
frequent scattering between the neutrinos and the more massive
nucleons and nuclei only allows for momentum transfer---at least, of
course, if we neglect for a moment the possibility of intrinsic
excitations of nuclei in neutrino scattering events (see section
\ref{sec:inelasticNN} for a discussion of that aspect).

Crucial for the total deleptonization during core collapse is the
physics in the density range \( 5\times 10^{10}-5\times 10^{12} \)
g/cm\( ^{3} \) inbetween the two extremes, free escape and trapping.
Figure \ref{fig:tauegydens.ps} visualizes the interplay between
neutrino production, neutrino thermalization, and neutrino diffusion
in a snapshot at bounce in model G15
\cite{Liebendoerfer.Rampp.ea:2005}.  Figure \ref{fig:tauegydens.ps} is
built around the electron chemical potential, \( \mu _{e} \) (thin
solid line), which sets the energy scale for reactions with the
degenerate electrons. Also shown is the quantity \( \mu _{e}-\mu
_{n}+\mu _{p} \) (dotted line) where \( \mu _{n}-\mu _{p} \) is the
difference between the neutron and proton chemical potentials.  The
quantity represents the neutrino chemical potential in regions where
the neutrinos are in equilibrium with the fluid. Thick solid lines
running from the upper left corner to the lower right indicate the
optical depth, \( \tau \), as a function of density and neutrino
energy. Contour lines for \( \tau =1 \) (in average one scattering
before escape), \( \tau =10 \) (in average 10 scatterings before
escape, etc.), and \( \tau =100 \) are shown. The contour lines
approximately run along \( E^{2}\rho ^{2/3}=const. \) because of the
increase of the scattering cross sections proportional to the squared
neutrino energy \cite{Bethe:1990}. The asterisks represent the average
energy of emitted neutrinos by electron capture on free protons and on
nuclei. For each asterisk in the graph, the trajectory of the most
probable escape in the two-dimensional space of neutrino energy and
matter density has been calculated and drawn with a thin solid line.
Note that the trajectories only connect the most probable of all
possible displacements of test neutrinos in the density-energy space.
They cannot represent a rate of actual displacements.  The
trajectories immediately confirm that neutrinos on the left hand side
of the \( \tau =1 \) line escape without further energy change while
the neutrinos on the right hand side of the \( \tau =100 \) line tend
to scatter to the neutrino Fermi energy on a much smaller time scale
than for the diffusion in space. The holes they leave in the
equilibrium distribution will immediately be refilled by new neutrino
emissions or scatterings.  Note also the trajectory from the
additional \( 5 \) MeV test particle at \( 2\times 10^{13} \) g/cm\(
^{3} \), which can only scatter to the Fermi surface because the
neutrinos are degenerate (The small steps in the Fermi energy are of
numerical nature and due to the finite resolution of the energy phase
space in the multi-group neutrino treatment).  The other two \( 5 \)
MeV test particles at \( 5\times 10^{12} \) and \( 10^{13} \) g/cm\(
^{3} \) demonstrate that at these densities very low energy neutrinos
start to escape due to the relatively low scattering cross sections at
these energies \cite{Mazurek:1973}. This frees phase space that is
replenished from high energy neutrinos that are now able to
down-scatter as indicated by the trajectories in the interesting
density range between \( 5\times 10^{10} \) and \( 5\times 10^{12} \)
g/cm\( ^{3} \). High energy neutrinos emitted at \( 10^{12} \) g/cm\(
^{3} \) finally escape with a lower energy than neutrinos emitted at a
lower energy at \( 10^{11} \) g/cm\( ^{3} \).  Hence, the most
significant deleptonization relies on a two-dimensional diffusion
process in the space spanned by neutrino energy and matter density and
takes place during the transition from the density where neutrinos
stream freely to the density where they are trapped.

The physics of the core collapse does not depend on the evolution of
the electron fraction alone. Also the entropy plays an important role.
In general the entropy is low during the collapse as most of the
energy is rather stored in internal excitation energy of the nuclei
than in thermal energy of the gas due to the large partition functions
for nuclei~\cite{Bethe.Brown.ea:1979}. However, the entropy is changed
by weak interactions in the core and has important consequences for
the evolution (e.g., an increase in entropy will increase the number
of protons available for electron capture).  The entropy change \(
\Delta s \) for a deleptonization \( \Delta Y_{e} \) is given in Ref.
\cite{Bruenn:1985} and references therein,
\begin{equation}
\label{eq:entropy.change.2}
T\Delta s=-\Delta Y_{e}\left( \mu _{e}-\mu _{n}+\mu _{p}\right) +\Delta q,
\end{equation}
where \( \Delta q \) is the energy transfer between matter and
neutrinos.  If we are not too strict about where in space the energy transfer
between matter and neutrinos takes place (for neutrino-electron
scattering rather along one of the trajectories in Fig.
\ref{fig:tauegydens.ps} than in one and the same fluid element), one may
simply set \( \Delta q=\Delta Y_{e}E_{\nu }^{\text{escape}} \) where
\( E_{\nu }^{\text{escape}} \) is the average energy of the neutrinos
when they escape the star \cite{Bethe:1990}. This leads to the following
expression for the entropy change (\ref{eq:entropy.change.2}):
\begin{equation}
\label{eq:entropy.change}
T\Delta s=-\Delta Y_{e}\left( \mu _{e}-\mu _{n}+\mu _{p}
-E_{\nu}^{\text{escape}}\right).
\end{equation}

Figure \ref{fig:tauegydens.ps} now illustrates, that up to a density
of \( 2\times 10^{11} \) g/cm\( ^{3} \) we find \( E_{\nu
}^{\text{escape}}>\mu _{e}-\mu _{n}+\mu _{p} \).  The entropy in this
regime is decreasing. Therefore, also the entropy might slightly
contribute to the convergence to a "norm" trajectory during the
deleptonization: If a fluid element in one model has a higher
temperature than in another model, more free protons are available for
electron capture and the fluid element cools more rapidly toward the
temperature in the other model. We also note that in this regime
electron capture on nuclei produces neutrinos with lower escape
energies and therefore contributes less to an entropy decrease than
electron capture on free protons \cite{Bruenn.Haxton:1991}.  Between
\( \rho =2\times 10^{11} \) g/cm\( ^{3} \) and \( \rho =2\times
10^{12} \) g/cm\( ^{3} \), Fig.  \ref{fig:tauegydens.ps} indicates
that \( E_{\nu }^{\text{escape}}<\mu _{e}-\mu _{n}+\mu _{p} \) such
that the entropy increases because a part of the locally emitted
neutrino energy is returned to the fluid by the thermalization of the
neutrino. At even higher densities, only the thermalization down to
the neutrino chemical potential occurs on a fast time scale such that
fast electron fraction changes do no longer cause entropy changes
because of \( \mu _{\nu }=\mu _{e}-\mu _{n}+\mu _{p} \). The energy
difference between \( \mu _{\nu } \) and \( E_{\nu }^{\text{escape}}
\) is only returned to the fluid on the much longer diffusion time
scale relevant for changes of the lepton fraction \( \Delta Y_{\ell }
\).  These expectations are consistent with the actual entropy
evolution at the center shown in Fig. \ref{fig:evol}.

\subsection{Electron capture rates}
\label{sec:electr-capt-during}

\begin{figure}[htb]
  \centering
  \includegraphics[angle=270,width=0.9\linewidth]{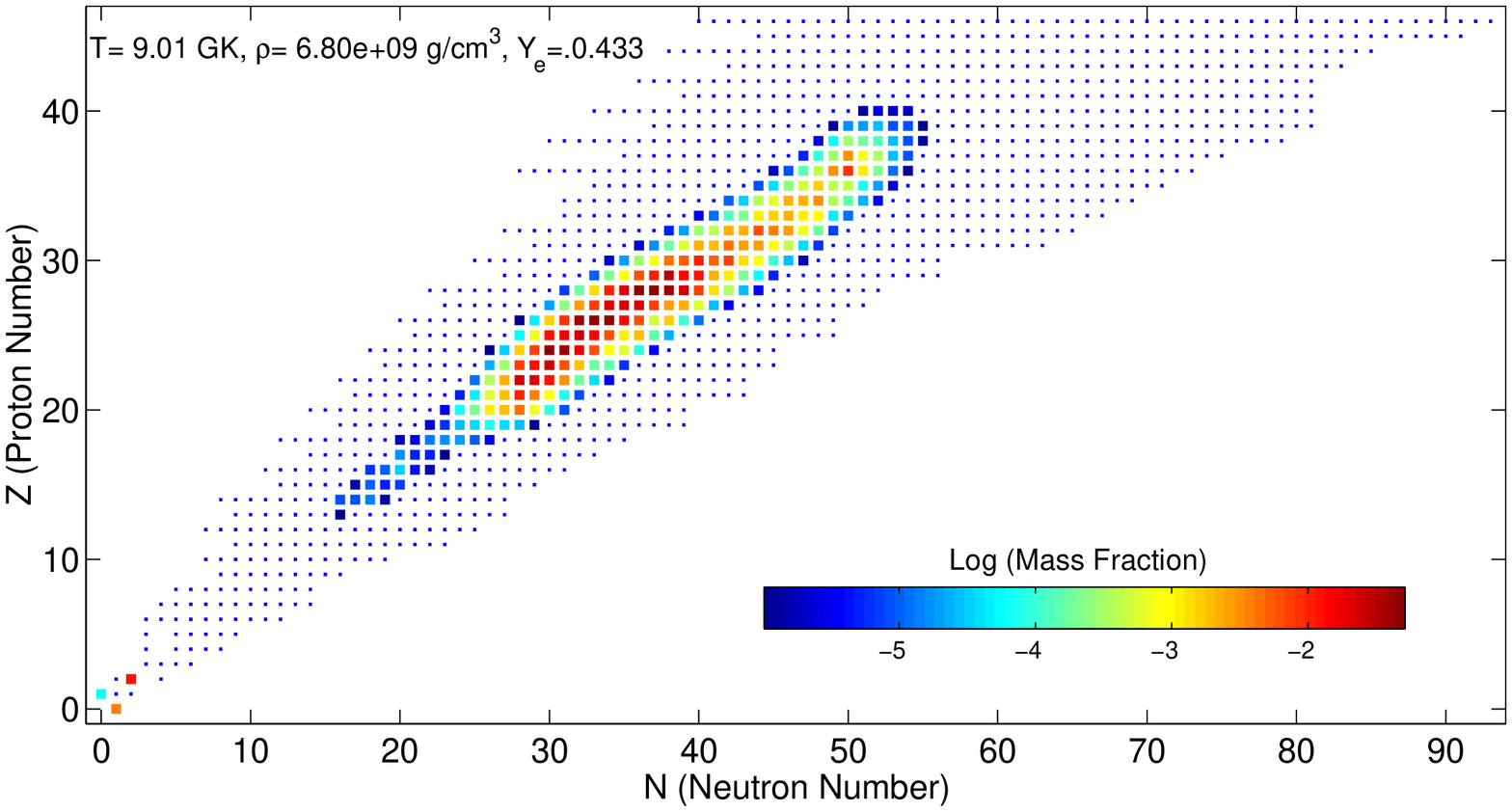}\\
  \includegraphics[angle=270,width=0.9\linewidth]{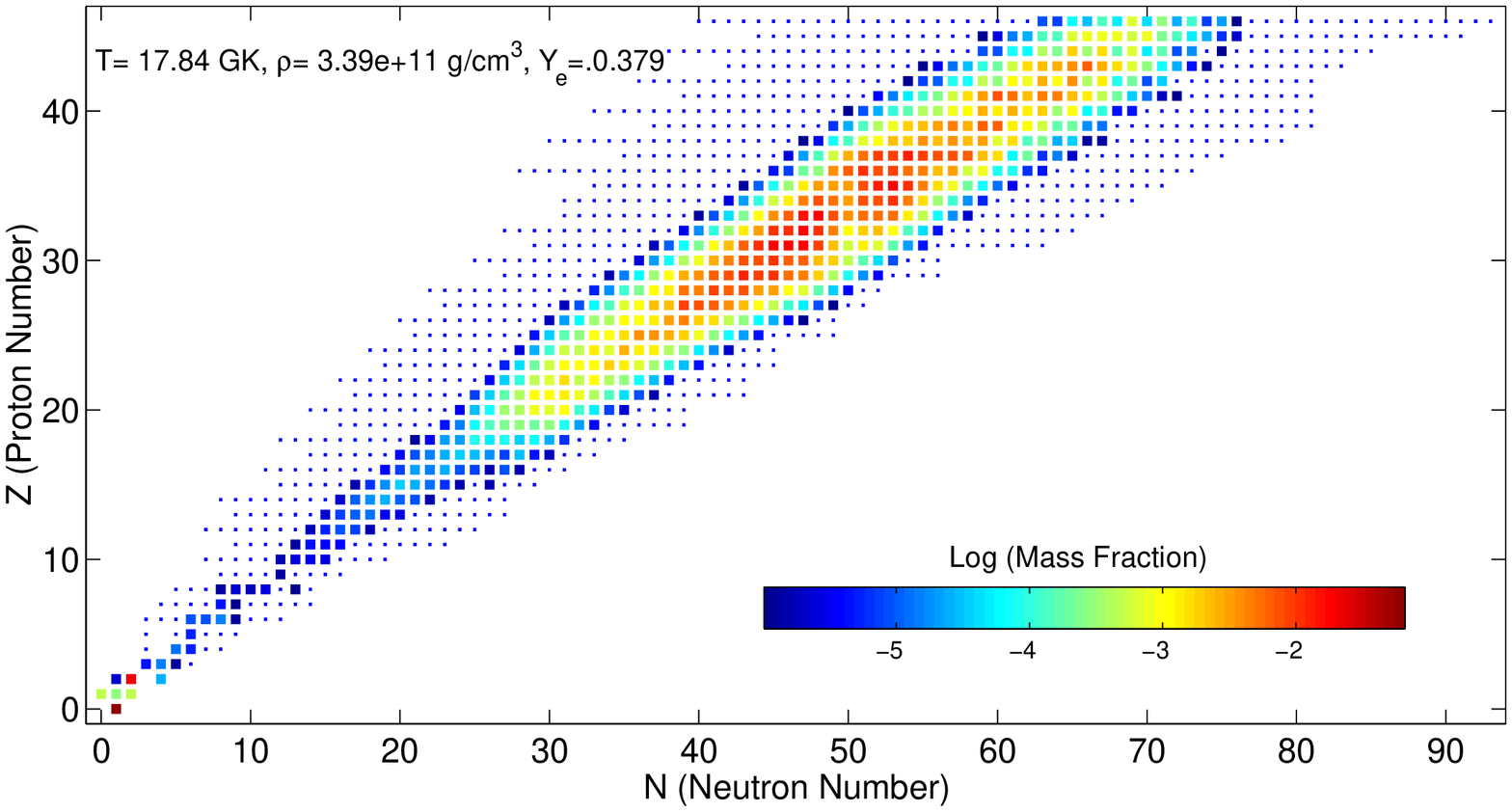}\\
  \caption{Abundances of nuclei for two set of conditions during the
    collapse. The upper panel represents typical conditions during
    the early collapse while the lower panel shows conditions near
    to neutrino trapping. A Saha-like NSE code has been used in the
    calculation of the abundances~\cite{Hix:1995} (Adapted
    from~\cite{Hix.Mezzacappa.ea:2003}).\label{fig:abundances}}
\end{figure}

Calculations of the reaction rate for electron capture in the
collapsing core requires two components: the appropriate electron
capture reaction rates and the knowledge of the nuclear composition.
The coupling of electron capture rates to energy-dependent neutrino
transport adds an additional requirement: information about the
spectra of emitted neutrinos. These spectra can be parametrized using
the prescription of Ref.~\cite{Langanke.Martinez-Pinedo.Sampaio:2001}.
During the collapse most of the collapsing matter survives in heavy
nuclei as the entropy is rather low~\cite{Bethe.Brown.ea:1979}. $Y_e$
decreases during the collapse due to electron capture, making the
matter composition more neutron rich and hence favoring increasingly
heavy nuclei. Unlike in stellar evolution and supernova
nucleosynthesis simulations, where the nuclear composition is tracked
in detail via a reaction
network~\cite{Woosley:1986,Hix.Thielemann:1999}, the composition used
in supernova simulations is calculated by the equation of state, which
assumes nuclear statistical equilibrium (NSE). Typically, the
information about the nuclear composition provided by the equation of
state is limited to the mass fractions of free neutrons and protons,
$\alpha$ particles and the sum of all heavy nuclei, as well as the
identity of an average heavy nucleus, calculated either in the liquid
drop framework~\cite{Lattimer.Swesty:1991} or based on a relativistic
mean field model~\cite{Shen.Toki.ea:1998a,Shen.Toki.ea:1998b}. It
should be noted that the most abundant nucleus is not necessarily the
nucleus which dominate electron capture during the infall phase.  For
the evaluation of reaction rates on nuclei, due to the dependence on
nuclear structure effects, a single nucleus approximation is not
sufficient. It must be replaced by an ensemble average.
Figure~\ref{fig:abundances} shows the typical nuclear
abundances~\cite{Hix.Mezzacappa.ea:2003} at two different conditions
taken from a collapse trajectory from
Ref.~\cite{Liebendoerfer.Messer.ea:2001}. Before discussing how the
detailed abundance information can be implemented in collapse
simulations we briefly review the foundation for the calculation of
electron capture rates during the infall phase.

Traditionally, in collapse simulations the treatment of electron
capture on nuclei is schematic and rather simplistic. The nuclear
structure required to derive the capture rate is then described solely
on the basis of an independent-particle model for iron-range nuclei,
i.e., considering only Gamow-Teller transitions from $f_{7/2}$ protons
to $f_{5/2}$ neutrons~\cite{Bethe.Brown.ea:1979,%
  Bruenn:1985,Mezzacappa.Bruenn:1993a,Mezzacappa.Bruenn:1993b}. In
particular, this model predicts that electron capture vanishes for
nuclei with neutron number $N\geq 40$, arguing that Gamow-Teller
transitions are blocked due to the Pauli principle, as all possible
final neutron orbitals are already occupied in nuclei with $N \geq
40$~\cite{Fuller:1982}. These nuclei dominate the composition for
densities larger than a few $10^{10}$~g~cm$^3$. As a consequence of
the model applied in previous collapse simulations, electron capture
on nuclei ceases at these densities and the capture is entirely due to
free protons. It has been pointed out~\cite{Cooperstein.Wambach:1984}
that this picture is too simple and that the blocking of the
Gamow-Teller transitions will be overcome by thermal excitations which
either moves protons into the $g_{9/2}$ orbit or removes neutrons from
the $pf$-shell, in both ways unblocking the GT transitions. According
to the work of Ref.~\cite{Cooperstein.Wambach:1984}, due to ``thermal
unblocking'' GT transitions dominate again for temperatures of the
order of 1.5~MeV. A more important unblocking effect, which is already
relevant at lower temperatures is expected from the residual
interaction which will mix the $g_{9/2}$ and higher orbitals with the
$pf$
shell~\cite{Langanke.Kolbe.Dean:2001,Langanke.Martinez-Pinedo.ea:2003}.

The calculation of electron capture on nuclei during the collapse
phase requires a model that is able to describe the correlations and
at the same time the high density of levels that can be thermally
populated at moderate excitation energies. Direct shell-model
diagonalizations are not yet possible due to the large model spaces
involved. The calculations can be done using the Shell Model Monte
Carlo approach~\cite{Koonin.Dean.Langanke:1997} which allows for the
calculation of nuclear properties at finite temperature in
unprecedentedly large model spaces. This model complemented with
Random Phase Approximation calculations for the computation of the
transitions necessary for the determination of the electron capture
rate has been used recently for the calculation of the relevant rates
for nuclei in the mass range
$A=65$--112~\cite{Langanke.Martinez-Pinedo.ea:2003}. Currently
electron capture rates are available for all the nuclei in the mass
range $A=17$--112 and for the temperatures and densities relevant for
the collapse~\cite{Fuller.Fowler.Newman:1982a,Oda.Hino.ea:1994,%
Langanke.Martinez-Pinedo:2001,Langanke.Martinez-Pinedo.ea:2003,%
Pruet.Fuller:2003}.

\begin{figure}[htbp]
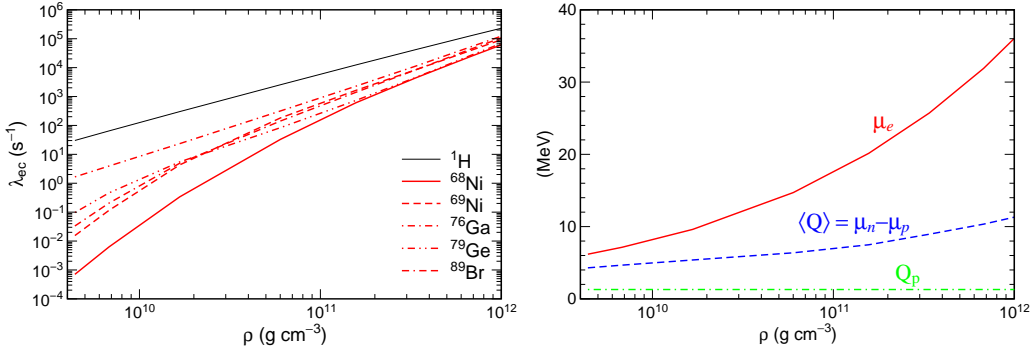

  \centering
  \includegraphics[bb=71 61 704 487,width=0.49\textwidth]{nucleirho.eps}%
  \includegraphics[bb=71 61 704 487,width=0.49\textwidth]{energies.eps}
  \caption{(left panel) Comparison of the electron capture rates on
    free protons and selected nuclei as function of density chemical
    potential along a stellar collapse trajectory taken
    from~\cite{Mezzacappa.Liebendoerfer.ea:2001}. (right panel) energy
    scales relevant for the determination of the electron capture
    rates. $\mu_e$ is the electron chemical potential.  $\langle Q
    \rangle$ and $Q_p$ are the average $Q$ value for electron capture
    on nuclei and protons respectively. \label{fig:ecrates}}
\end{figure}

Figure~\ref{fig:ecrates} compares the electron capture rates for free
protons and selected nuclei along a stellar trajectory taken
from~\cite{Mezzacappa.Liebendoerfer.ea:2001}. These nuclei are abundant
at different stages of the collapse. For all the nuclei, the rates are
dominated by GT transitions at low densities, while forbidden
transitions contribute sizably for $\rho \gtrsim
10^{11}$~g~cm$^{-3}$. The electron chemical potential $\mu_e$ and the
reaction $Q$ value are the two important energy scales of the capture
process. (They are shown on the right panel of
figure~\ref{fig:ecrates}.) The $Q$-value dependence of the capture
rate can be well approximated by~\cite{Fuller.Fowler.Newman:1985}: 

\begin{equation}
  \label{eq:lambda}
  \lambda = \frac{(\ln 2)B}{K} \left(\frac{T}{m_e c^2}\right)^5
  \left[F_4(\eta) + 2  
    \chi F_3(\eta) +
    \chi^2 F_2 (\eta)\right],
\end{equation}
where $\chi = Q/T$, $\eta=(\mu_e - Q)/T$, $K=6146$~s, and $B$
represents a typical (Gamow-Teller plus forbidden) matrix element. The
quantities $F_k$ are the relativistic Fermi integrals of order
$k$\footnote{The fermi integrals are defined by:
  $F_k(\eta)=\int_0^\infty dx x^k/(1+\exp(x-\eta)) $}.  $Q$ represents
the effective $Q$-value for the capture rate including thermal
effects. In equation~\eqref{eq:lambda}, we use the convention that $Q$
is positive for capture on neutron rich nuclei and protons.  For low
densities ($\lesssim 10^{10}$~g~cm$^{-3}$), $\mu_e \approx Q$ and the
term $F_2$ dominates. The rate is then larger for the nuclei with
smaller $Q$ values. For intermediate densities
($10^{10}$--$10^{11}$~g~cm$^{-3}$), the term $F_3$ dominates and the
rate still has some dependence on the $Q$-value. For high densities
($\gtrsim 10^{11}$~g~cm$^{-3}$), $\mu_e \gg Q$ so that the term $F_4$
dominates and the rate becomes independent of the $Q$-value, depending
only on the total GT strength, but not its detailed distribution.

According to figure~\ref{fig:ecrates}, the electron capture rate on a
proton is larger than that for individual nuclei. However, this is
misleading as the low entropy keeps the protons significantly less
abundant than heavy nuclei during the collapse. Simulations of core
collapse require then the knowledge of the detailed abundances of all
the nuclei present. As the commonly used equations of
state~\cite{Lattimer.Swesty:1991,Shen.Toki.ea:1998a,Shen.Toki.ea:1998b}
do not provide this detailed information, a Saha-like NSE was used for
the calculation of the abundances in
refs.~\cite{Langanke.Martinez-Pinedo.ea:2003,Hix.Messer.ea:2003}. Once
the abundances are considered the reaction rate for electron capture
on heavy nuclei ($R_h = \sum_i Y_i \lambda_i$, where the sum runs over
all the nuclei present and $Y_i$ denotes the number abundance of
species $i$) dominates over the one of protons ($R_p=Y_p \lambda_p$)
by roughly an order of magnitude throughout the
collapse~\cite{Langanke.Martinez-Pinedo.ea:2003,Hix.Messer.ea:2003}.
The first simulations that considered electron capture on nuclei with
$N\ge40$ were probably the ones of Ref.~\cite{Bruenn.Haxton:1991}.
These authors determined the electron-capture rate by detailed balance
from their computed $\nu_e$ absorption rate. This corresponds to
electron capture on the thermally populated Fermi and GT$_-$
resonances in the parent nucleus shown in figure~\ref{fig:scheme}
(also called backresonances according to the notation
of~\cite{Fuller.Fowler.Newman:1980}) that decay to the daughter ground
state. This backresonance contribution is in general, depending on the
nucleus, a factor between 10 and 100 smaller than the direct
transition from the parent ground state to the daughter Gamow-Teller
resonance. This explains why, in contrary to the results of
refs.~\cite{Langanke.Martinez-Pinedo.ea:2003,Hix.Messer.ea:2003},
electron capture on nuclei was too small in
Ref.~\cite{Bruenn.Haxton:1991} so that it was concluded that electron
capture on protons dominates over capture on nuclei.

\begin{figure}[htbp]
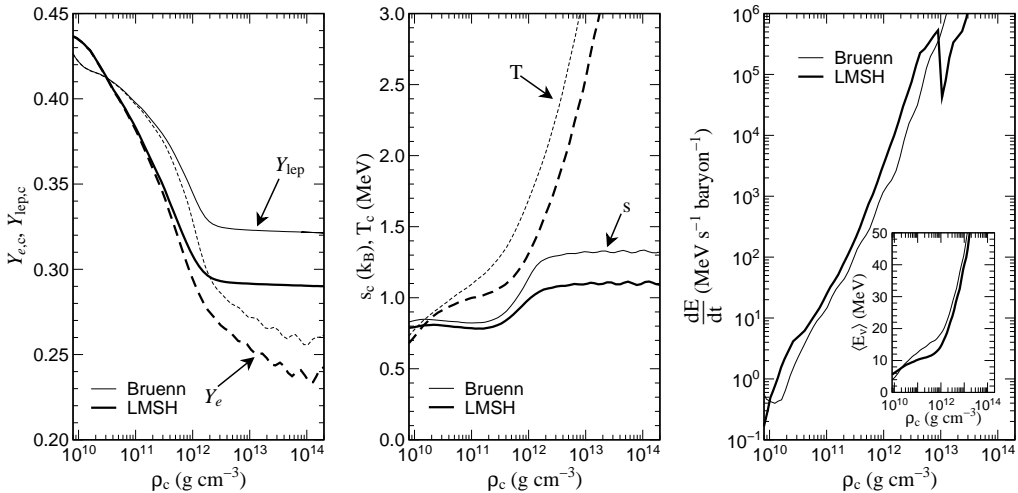

  \centering
  \includegraphics[bb=52 117 449 696,width=0.32\textwidth]{martinez_fig2a.eps}%
  \hspace{0.001\textwidth}%
  \includegraphics[bb=52 117 449 696,width=0.32\textwidth]{martinez_fig2b.eps}%
  \hspace{0.019\textwidth}%
  \includegraphics[bb=52 117 449 696,width=0.32\textwidth]{martinez_fig2c.eps}
  \caption{Comparison of the evolution of several quantities at the
    center of a 15~M$_\odot$: $Y_e$ is the number of electrons per
    baryon, $Y_{\text{lep}}$ is the number of leptons per baryon $s$
    is the entropy per baryon, $T$ is the temperature, $dE/dt$ is the
    neutrino energy emission rate per baryon, $\langle E_\nu\rangle$
    is the average energy of the emitted neutrinos.  An abrupt change
    in the emission occurs when the domain where the LMSH rates are
    specified is exceeded at a density of \( 10^{13} \) g/cm\(^3\).
    The initial presupernova model was taken
    from~\cite{Heger.Langanke.ea:2001}. The thin line is a simulation
    using the Bruenn parametrization~\cite{Bruenn:1985} while the
    thick line uses the LMSH rate set (see text). Both models include
    the general relativistic effects discussed in
    section~\ref{sec:hydrodynamics} and were calculated by the
    Garching collaboration (Courtesy of M. Rampp and H.-Th.
    Janka).\label{fig:evol}}
\end{figure}

The effects of the more realistic implementation of electron capture
on heavy nuclei have been evaluated as described above in independent
self-consistent neutrino radiation hydrodynamics simulations by the
Oak Ridge and Garching
collaborations~\cite{Hix.Messer.ea:2003,Rampp.Janka.Private}.  The
results obtained by the Garching collaboration are shown in
Fig.~\ref{fig:evol}~\cite{Rampp.Janka.Private} for the evolution of
several quantities at the center of a 15~M$_\odot$ star using the
standard treatment of Bruenn~\cite{Bruenn:1985} and the new rates for
heavy nuclei (denoted LMSH).  With the improved treatment of electron
capture rate on heavy nuclei the total electron capture rate (heavy
nuclei plus protons) is larger than in the Bruenn treatment resulting
in smaller \( Y_e \) and \( Y_{\text{lepton}} \) values. The left
panel shows the evolution of these quantities at the center of the
star as functions of the increasing density during collapse. The small
oscillations in the trajectories above \( 5\times10^{12} \) g/cm\( ^3
\) are due to the same numerical effect of finite energy resolution
under neutrino-degenerate conditions that we have already seen in
Fig.~\ref{fig:tauegydens.ps}.  The panel demonstrates that the Bruenn
treatment leads to a faster deleptonization for representative nuclei
with \( N<40 \) (at low density) while this channel is blocked
afterwards such that the further deleptonization is delayed and falls
short of the LMSH treatment where electron captures on nuclei continue.
The panel on the right shows the energy per baryon and second emitted
in neutrinos at the center of the star. The inset shows the mean
energy of the produced neutrinos. With the LMSH treatment, more
neutrinos are produced with lower mean energy than with the Bruenn
treatment. The center panel shows the entropy and temperature
evolution at the center of the star. The general entropy trajectory is
consistent with the discussion following Eq.
(\ref{eq:entropy.change}). The entropy is slowly descending at low
density because \( E_{\nu}^{\text{escape}} \) is larger than \( \mu_e
- \mu_n + \mu_p \). As soon as neutrino-electron scattering becomes
effective to downscatter neutrinos, \( E_{\nu}^{\text{escape}} \)
falls below above chemical potential difference (see Fig.
\ref{fig:tauegydens.ps}, left panel) and the transferred heat starts
to increase the fluid entropy until the diffusion time scale becomes
larger than the dynamical time scale.  With the LMSH treatment one
would now be tempted to expect that the central entropy would rise to
larger values than with the Bruenn treatment because the
deleptonization \( \Delta Y_e \) in Eq. (\ref{eq:entropy.change})
increases according to the values shown in the left panel.
Additionally the neutrinos are emitted with lower energies so that \(
E_{\nu}^{\text{escape}} \) might also decrease somewhat to favor a
steep entropy increase.  However, the center panel shows clearly that
the LMSH treatment leads to smaller central entropies. In order to
better illustrate speculations about the possible reasons of this
interesting result, we performed a simple test calculation for a
comparison in Fig. \ref{fig:tauegydens.ps} where one state in the
neutron \( f_{5/2} \) orbit was kept unblocked in the Bruenn
treatment.  One immediately sees in the right panel that the average
energy of emitted neutrinos (asterisks) is lower than with the Bruenn
treatment and that less neutrino-electron scattering takes place
before the escape.  This difference in the thermalization path reduces
the diffusion time. For an understanding of the entropy differences
however, it is crucial to realize that the difference in the
deleptonization between the Bruenn and LMSH rates occurs {\it before}
the density is reached where the entropy rises. We define a first
phase, reaching from low densities to \( \rho=3\times 10^{10} \)
g/cm\(^3\), so that the net deleptonization for the two rates is
exactly the same (left panel in Fig. \ref{fig:evol}).  As both rates
include electron capture on nuclei at these densities, Eq.
(\ref{eq:entropy.change}) does not predict noticeable differences in
the entropy evolution.  In a second phase, reaching from \(
\rho=3\times 10^{10} \) to \( \rho=2\times 10^{11} \) g/cm\(^3\),
electron capture on nuclei is blocked in the Bruenn treatment and a
significantly smaller net deleptonization \( -\Delta Y_e \) in Eq.
(\ref{eq:entropy.change}) occurs than with the LMSH rates. However,
Fig. \ref{fig:tauegydens.ps} shows that in this phase the second
factor, \( \mu_e - \mu_n + \mu_p - E_{\nu}^{\text{escape}} \), is much
smaller than in the LMSH case because electron capture on free protons
produces neutrinos with larger escape energies \(
E_{\nu}^{\text{escape}} \) than electron capture on nuclei. The two
effects cancel in the evaluation of \( T\Delta s\) in Eq.
(\ref{eq:entropy.change}) and the entropy evolution in this phase is
again similar with both sets of rates. Now, in the third phase,
reaching from \( \rho=2\times 10^{11} \) to \( \rho=2\times 10^{12} \)
g/cm\(^3\), there is no significant difference in the net
deleptonization for the two cases anymore, \( -\Delta Y_e\sim 0.09 \),
but the lower \( Y_e \) values increase the difference
between the neutron and proton chemical potentials, \( \mu_n - \mu_p
\), in the LMSH case.  On the other hand, the energy of escaping
neutrinos, \( E_{\nu}^{\text{escape}} \), decreases, but only about
half as much. Both effects together cause a difference \( \sim 2 \)
MeV in the factor \( \mu_e - \mu_n + \mu_p - E_{\nu}^{\text{escape}}
\), which seems consistent with the displayed difference in the
entropy evolution before trapping in the center panel of Fig.
\ref{fig:evol}.  The increased neutron to proton asymmetry in the LMSH
case also leads to heavier nuclei with higher opacities for a given
density and neutrino energy (see section \ref{sec:opacity}). This
shifts the deleptonization region to slightly lower densities where
the energy scales are smaller. This might also contribute to the
entropy difference.

\subsection{Inelastic Neutrino-Nucleus Interactions}
\label{sec:inelasticNN}

Although the classical thermalization mechanism for the neutrinos
proceeds through inelastic scattering on electrons,
Eq.~\eqref{eq:nuenue}, neutrino-induced reactions on nuclei can also
occur~\cite{Haxton:1988,Bruenn.Haxton:1991}. Charged-current
$(\nu_e,e^-)$ reactions on nuclei are unimportant during the collapse
compared with inelastic neutrino-electron scattering due to the strong
Pauli blocking of the final electron phase
space~\cite{Bruenn.Haxton:1991}. This is so even if finite temperature
effects are considered~\cite{Sampaio.Langanke.Martinez-Pinedo:2001}.

However, according to Ref.~\cite{Bruenn.Haxton:1991}, neutrino-nucleus
inelastic scattering plays an as important role as neutrino-electron
scattering. The influence of finite temperature was studied in
Ref.~\cite{Fuller.Meyer:1991} using an independent particle model.
More recently, neutrino cross sections for finite temperature have
been calculated using shell-model GT
distributions~\cite{Sampaio.Langanke.ea:2002,Juodagalvis.Langanke.ea:2004}.
This approach has been validated for the calculation of
neutrino-nucleus inelastic scattering against high resolution magnetic
dipole data~\cite{Langanke.Martinez-Pinedo.ea:2004}. Finite
temperature enhances the cross section for low energy neutrinos. This
is due to the fact that at finite temperatures the initial nucleus can
reside in excited states which can be connected to the ground state by
large GT matrix elements. These states are deexcited by inelastic
neutrino scattering. In this case, contrarily to what happens in
neutrino-electron scattering, the final neutrino energy is larger than
the initial. Currently no hydrodynamical collapse simulations are
available that include both electron capture on nuclei and
neutrino-nucleus inelastic scattering.  Nevertheless, the simple fact
that the cross sections are of similar magnitude than scattering on
electrons~\cite{Sampaio.Langanke.ea:2002} suggest their inclusion in
supernova neutrino transport codes.

Other neutrino processes such as nuclear deexcitation by neutrino pair
production, Eq.~\eqref{eq:AAnunu}, have been discussed in
Ref.~\cite{Fuller.Meyer:1991}, but the estimated rates are probably
too small for these processes to be important during the collapse. 

\subsection{Neutrino opacities}
\label{sec:opacity}

After the detailed description of the new electron capture rates in
section \ref{sec:electr-capt-during} and the emphasis of the
importance of neutrino thermalization in sections
\ref{sec:deleptonization} and \ref{sec:inelasticNN}, we conclude the
discussion of nuclear input for core-collapse models with a few
pointers to the third relevant ingredient: neutrino opacities. The
fundamental neutrino opacity in core collapse is provided by neutrino
scattering on nucleons. Depending on the distribution of the nucleons
in space and the wavelength of the neutrinos, various important
coherence effects can occur: Most important during collapse is the
binding of nucleons in nuclei with a density contrast of several
orders of magnitude to the surrounding nucleon gas.  For elastic
neutrino-nucleus scattering one usually makes the simplifying
assumption that the nucleus has a $J=0^+$ spin/parity assignment. This
is appropriate for the ground state of even-even nuclei.  The
scattering process is then restricted to the Fermi part of the neutral
current (pure vector coupling)
\cite{Freedman:1974,Tubbs.Schramm:1975}.  Because of coherent
scattering, the cross section scales with $A^2$, except from a
correction $\sim (N-Z)/A$ arising from the neutron excess. This
assumption is, in principle, not correct for the ground states of
odd-$A$ and odd-odd nuclei and for all nuclei at finite temperature,
as then $J \ge 0$ and the cross section will also have an axial-vector
Gamow-Teller contribution.  However, the relevant GT$_0$ strength is
not concentrated in one state, but rather fragmented over many nuclear
levels. Thus, one can expect that the GT contributions to the elastic
neutrino-nucleus cross sections are in general small enough to be
neglected. A useful comparison of inverse mean free paths at the
important density \( \rho =10^{12} \) g/cm\( ^{3} \) is given in
\cite{Bruenn.Mezzacappa:1997}. It is found that \( \lambda _{\nu
  +n}/\lambda _{\nu +A}\sim 3\times 10^{-2} \), \( \lambda _{\nu
  +e}/\lambda _{\nu +A}\sim 2.5\times 10^{-2} \), \( \lambda _{\nu
  +He}/\lambda _{\nu +A}\sim 10^{-4} \), and \( \lambda _{\nu
  +p}/\lambda _{\nu +A}\sim 5\times 10^{-5} \).

Further corrections are necessary: with an increasing ratio between
the Coulomb potential of the positively charged ions and their thermal
energy, the average separation between nuclei will more strongly peak
around the value of most efficient packing. The neutrino opacities are
then to be corrected by an ion-ion correlation function \(
\left\langle S_{ion}\left( E_{\nu }\right) \right\rangle <1 \)
\cite{Itoh:1975,Itoh.Totsuji.ea:1979,Horowitz:1997}. Its consideration
in core collapse simulations lowers the trapped lepton fraction at
bounce by \( 0.015 \) and increases the central entropy per baryon by
\( 0.12 \) kB \cite{Bruenn.Mezzacappa:1997}. It is found that the
sizable entropy increase is not only due to the increased
deleptonization, but also to the fact that the correlation effect is
most pronounced for the low energy neutrinos with a long wavelength.
Hence the entropy increase is additionally supported by a decrease of
\( E_{\nu }^{\text{escape}} \) in the discussion following Eq.
(\ref{eq:entropy.change}). As current core collapse models proceed
toward including the full ensemble of nuclei instead of just a
representative nucleus (cf. Fig. \ref{fig:abundances}), it becomes
rather non-trivial how to adequately determine correlation effects in
the ion mixture, see for example
\cite{Itoh.Totsuji.ea:1979,Itoh.Hayashi.ea:1996}.

The situation is even more interesting in the phase transition from isolated
nuclei to bulk nuclear matter where the nuclei or the holes inbetween
them are strongly deformed. Various pasta-like shapes may be assumed.
Correlation effects in this phase could also affect the neutrino opacities
\cite{Horowitz.Perez-Garcia.Piekarewicz:2004}. For an immediate effect
on core collapse, however, it would be required that this {}``pasta-phase''
would reach to fairly low densities in order to affect the opacities
at the neutrinospheres where the neutrino luminosities and spectra
are set (cf. Fig. \ref{fig:tauegydens.ps}).

A more detailed description of the neutrino opacities in nuclear
matter is given elsewhere in this volume
\cite{Burrows.Reddy.Thompson:2004}. An extensive quantitative overview
of the rates of most reactions has been provided in
Ref.~\cite{Bruenn.Haxton:1991}. Useful fitformulae are collected in
Ref.~\cite{Itoh.Totsuji.ea:1979}.

\section{Effects on the postbounce evolution}
\label{sec:postbounce}

Stellar core collapse and supernova explosion have often been
discussed as one and the same event. This view has also
observationally been confirmed by the detection of few neutrinos from
supernova 1987A
\cite{Bionta.Blewitt.ea:1987,Hirata.Kajita.ea:1987,Alekseev.Alekseeva.ea:1988}.
The connection has further been nourished by the early theory of a
prompt explosion mechanism where, as a consequence of bounce, a strong
shock rushes through the outer core to eject the outer layers.
However, we are not aware of any simulation with reasonable neutrino
physics that would have predicted a prompt explosion since the
Lattimer-Swesty equation of state \cite{Lattimer.Swesty:1991} has
become standard.  In contrary, due to the stiff behaviour of the
equation of state around nuclear density the initial shock is weak as
discussed in section \ref{sec:hydrodynamics}. Additionally it starts at a very deep mass
coordinate because of the efficient deleptonization discussed in
sections \ref{sec:deleptonization}-\ref{sec:electr-capt-during}.
In fact, recent spherically symmetric simulations of the
postbounce evolution predict that the shock stalls already \( \sim 3
\) ms after bounce, i.e. before or at the time the neutrino burst is
launched from neutrino-transparent layers and \( \sim 40 \) ms
\emph{before} a gain layer can develop for neutrino heating to become
effective. Many publications of supernova models include figures with
shock trajectories that show a continued expansion of the shock front
to about \( 150 \) km at \( 100 \) ms after bounce. Note, however,
that there is no valid concept of shock energy that reaches beyond the
few milliseconds after which the shock has stalled. The continued
expansion of the accretion front over at least \( 90\% \) of the \(
\sim 100 \) ms interval is due to the accumulation of
material on the protoneutron star: heavy nuclei from the outer layers
are accelerated in the gravitational potential until they fall upon
the accretion front where their kinetic energy is converted into heat.
The accretion front is displaced to a larger radius mainly by the
increased volume of the accumulated hot and dissociated matter and
not by shock energy.

The sophisticated nuclear physics required in core collapse models
and the complicated dynamics in the ensuing supernova explosion suggest
a point of view where the theory of core collapse may well be as detached
from supernova explosions as, for example, the dynamics of star formation
from stellar evolution. On the one hand, core collapse involves the
nuclear and weak interaction physics of heavy nuclei under electron
degenerate conditions in layers that are unstable to gravitational
collapse. The supernova explosion, on the other hand, involves fundamentally
multi-dimensional neutrino-radiation-hydrodynamics in hot dissociated
matter that is in convective motion around a possibly magnetized
compact object. Also the technical refinement of simulations of the
collapse phase has always been roughly \( 10 \) years ahead of the
corresponding simulations of the postbounce phase. Based on the current
understanding, it would be an optimistic expectation that any very
relevant change in one domain (core collapse) would automatically
induce a revolution in the other (supernova explosion).

Important changes have been discovered in the core collapse event.  In
conventional simulations, electron capture on nuclei has been treated
as blocked for nuclei with \( N>40 \).  Recent models that include
realistic electron capture rates beyond this limit show that electron
capture on these nuclei are not only relevant, but even provide the
dominant channel for the deleptonization during
collapse~\cite{Langanke.Martinez-Pinedo.ea:2003,Hix.Messer.ea:2003},
with important consequences for core-bounce: In the interior of the
protoneutron star the electron fraction is reduced by \( \sim 10\% \)
when compared to models that don't include electron captures on
nuclei.  This change corresponds to a reduction of \( 20\% \) in the
mass of the homologous core and translates directly to a reduction of
the mass interior to the point of shock formation. Moreover, the shock
is launched with a \( \sim 15\% \) smaller velocity difference. The
weaker shock with more iron core overlying it turns even earlier (and
at a deeper mass coordinate) into an expanding accretion front than in
models with {}``standard'' physics. Furthermore, the improved
treatment of electron capture also affects the outer layers where the
average neutron number is well below \( N=40 \). For these conditions
the treatment of electron capture by Bruenn \cite{Bruenn:1985} results
in more electron capture than with the shell-model based electron
capture rates \cite{Langanke.Martinez-Pinedo:2000}.  The smaller
neutronization in the outer layers slows the collapse in these layers
which further diminishes the growth of the electron capture rate.
These changes delay the accretion of matter and reduce the ram
pressure opposing the expansion of the accretion front. Both effects
can modify the trajectory of the accretion front as a function of
time. 

In spherically symmetric simulations the accretion front reaches a
maximum expansion \( \sim 100 \) ms after bounce when the compression
of matter in the cooling region starts to exceed the volume of
accumulated shock- and neutrino-heated matter at the accretion front.
However, because the shock stalls within milliseconds in old and new
models, the significant changes in the core collapse physics cannot
dynamically push through to the explosion phase. Also the temporal
variations in the accretion rate tend to average over \( \sim 100 \)
ms, so that the maximum radius in the postbounce expansion of the
accretion front is not significantly sensitive to the improvements of
the electron capture rates on nuclei.  Nevertheless, core collapse
still sets the stage for a delayed supernova explosion and several
consequences of the nuclear physics during core collapse survive for a
longer time than the often overemphasized shock dynamics, or are in
principle directly observable.

Due to the extended accretion phase of delayed explosions,
most if not all of the iron core is packed
in the protoneutron star. Hence, differences in the physics of core
collapse will most likely translate to differences in the interior of
the protoneutron star. The deleptonization during collapse and the
early shock energetics leave a clear imprint on the entropy profile at
high densities. Indeed, models that include electron capture on nuclei
show considerably different entropy and lepton fraction gradients when
compared to standard treatments of electron capture
\cite{Hix.Messer.ea:2003}. These gradients are responsible for the
location, extent and strength of the protoneutron star convection and
other possible fluid instabilities. Seminal simulations by
\cite{Wilson:1985} favored a strong coupling between the protoneutron
star interior and the exterior layers subject to neutrino heating. The
equation of state used at that time suggested fluid instabilities
which, implemented by a mixing-length approach in spherically
symmetric simulations, supported the existence of delayed
neutrino-driven supernova explosions.  Analytical investigations based
on the Lattimer-Swesty equation of state \cite{Bruenn.Dineva:1996} and
recent numerical simulations \cite{Buras.Rampp.ea:2003} find
convectively stable neutrinospheres that efficiently block a rapid
release of neutrinos from the convectively unstable protoneutron star
interior. But a change in the nuclear input physics always carries the
potential to affect the structure of the protoneutron star in a way
that possibly revives the strong coupling of its interior to the outer
layers. A recent detailed investigation of the protoneutron star
stability is given in Ref. \cite{Bruenn.Raley.Mezzacappa:2004}.

\begin{figure}[htbp]
  \centering
  \includegraphics[height=0.6\textwidth,width=0.6\textwidth]{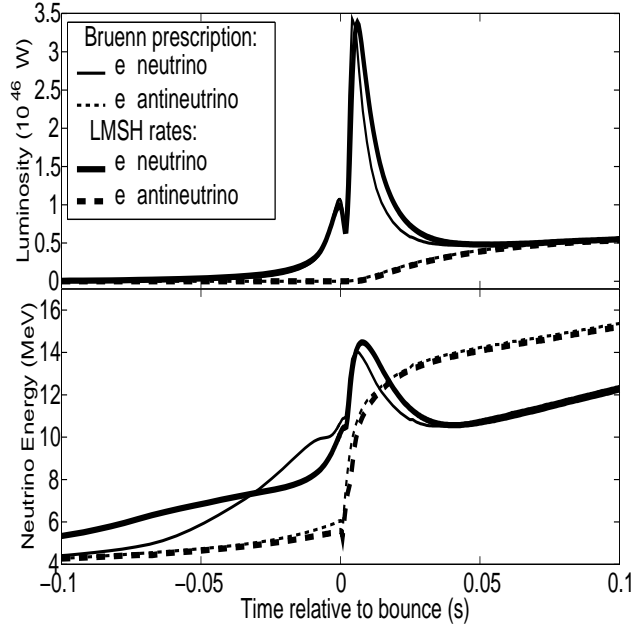}
  \caption{The neutrino luminosity and root-mean-square energy (at
    500~km) as a function of time from bounce for a 15 M$_\odot$
    model. The thin lines show this evolution for a simulation using the
    Bruenn parametrization~\cite{Bruenn:1985}, while the thick lines
    show this evolution for a simulation using the LMSH
    rates~\cite{Langanke.Martinez-Pinedo.ea:2003,Hix.Messer.ea:2003}. 
    The solid lines correspond to electron type neutrinos; the dashed
    lines correspond to electron type antineutrinos. Both cases were
    calculated in a general relativistic model
    by the Oak-Ridge collaboration (adapted
    from~\cite{Hix.Messer.ea:2003}). 
    \label{fig:nutime}}
\end{figure}

The neutrino luminosities and energies during collapse, bounce, and
the interesting time of the launch of the electron neutrino burst are
in principle more directly observable. Figure~\ref{fig:nutime} shows
the luminosity and root-mean-square energy of the emitted electron
neutrinos and antineutrinos (sampled at \( 500 \) km radius in the
comoving coordinate frame) between \( 100 \) ms before bounce and \(
100 \) ms after bounce. One model (thick lines) includes electron
captures on nuclei (LMSH) and the other (thin lines) uses the standard
Bruenn treatment. The luminosity shows a clear delay (\( 2 \) ms) in
the breakout burst due to a deeper launch of the shock in the LMSH
case. Consistent with the dependence of the neutrino burst intensity
on the shock strength \cite{Liebendoerfer.Rampp.ea:2005}, the LMSH
neutrino burst is also more extended in time. Before bounce the
neutrino luminosities are similar but the average energies are smaller
in the LMSH case due to the larger Q-values involved for electron
capture on nuclei as discussed above. After bounce, during the first
\( 50 \) ms the LMSH model emits \( \sim 15\% \) more energy than the
Bruenn model. This is mainly the result of a larger mean electron
neutrino energy in the LMSH model due to a deeper and hotter
neutrinosphere \cite{Hix.Messer.ea:2003}.

Neutrino-induced reactions on nuclei are expected to contribute only
modestly to the shock revival, due to the small abundance of nuclei
during this phase. It has been suggested that the shock revival can be
aided by ``preheating'' of matter ahead of the shock by absorption of
electron neutrinos emitted in the breakout burst~\cite{Haxton:1988}.
These neutrinos can partly dissociate the matter, mainly iron and
silicon isotopes, before the arrival of the shock. The early
explorations of Ref.~\cite{Bruenn.Haxton:1991} found no significant
preheating of unshocked matter. Once the matter is processed by the
shock, large amounts of $^4$He are present just below the stagnated
shock. The possibility of reheating this material by inelastic
$\nu_x$-$^4$He scattering, where $\nu_x$ denotes muon or tau neutrino
and antineutrino, or heating the material ahead of the shock by
$\nu_x$-nucleus inelastic scattering was also investigated in
Ref.~\cite{Bruenn.Haxton:1991} with no major effect
found. Nevertheless, only recently reliable neutrino-nucleus cross
sections have become available~\cite{Juodagalvis.Langanke.ea:2004}
that allow for a complete study of the influence of neutrino-nucleus
scattering both in the collapse and postbounce evolution. 

\section{Conclusions}

Core-collapse supernovae present a rich interplay between nuclear
physics and neutrino radiation (magneto-)hydrodynamics. Nuclear
physics sets the scene in which different hydrodynamical instabilities
could develop that eventually may produce robust explosions by the
neutrino-driven mechanism (see for example refs.
\cite{Burrows.Walder.ea:2004,Mezzacappa:2004,Janka.Buras.ea:2004} for
recent overviews over the activities of collaborations engaging in
accurate neutrino transport). The nuclear equation of state governs
stellar core collapse, core bounce and neutron star formation.  The
complicated interplay between deleptonization, neutrino
thermalization, and diffusion based on different weak interaction
processes---the most important of them is electron capture on
nuclei---determine the collapse dynamics, the position of shock
formation and the structure of the layers into which the shock
expands.  Advances in the calculations of these processes coupled to
improvements in one and multi-dimensional accurate neutrino transport
simulations has allowed to gain further insight to the physics of the
collapsing core. Future advances on the nuclear physics side should be
guided by the advent of new radiative-ion beam facilities that will
help to constrain the nuclear models used for the determination of the
equation of state and the different nuclear weak-interaction
processes.  These advances coupled to current efforts to develop
multidimensional magneto-hydrodynamical codes with accurate neutrino
transport will foster a definitive and quantitative understanding of
supernova explosions.

\end{document}